\documentclass{aa}

\usepackage{graphicx}
\usepackage{natbib}
\usepackage{txfonts}
\usepackage{hyperref}
\graphicspath{ {Figs/} }

\begin{document}

\title{Study of the bipolar jet of the YSO Th~28 with VLT/SINFONI: Jet morphology and H$_2$ emission. \thanks{Based on observations collected at the European Organisation for Astronomical Research in the Southern Hemisphere under ESO program 095.C-0892(A).}
}

\author{S. Melnikov \inst{1,2} \and P.~A. Boley \inst{3,4}\thanks{Former affiliation.} \and N.~S. Nikonova\inst{4}$^{\star{}\star{}}$ \and A. Caratti o Garatti \inst{5,6} \and
	R. Garcia Lopez \inst{6,5} \and B.~Stecklum \inst{1} \and J.~Eisl\"offel \inst{1} \and G. Weigelt \inst{7}}

\institute{Th\"uringer Landessternwarte Tautenburg, Sternwarte 5, D-07778 Tautenburg, Germany \and 
	National University of Uzbekistan, Physics Faculty, Department of Astronomy and Atmospheric Physics, 100174 Tashkent, Uzbekistan \and
	Visiting astronomer, Laboratoire Lagrange, Universit\'e C\^ote d\textquoteright Azur, Observatoire de la C\^ote d\textquoteright Azur, CNRS, Boulevard de l\textquoteright Observatoire, CS 34229, 06304, Nice Cedex 4, France \and
	Moscow Institute of Physics and Technology, 9 Institutskiy per., 141701 Dolgoprudny, Moscow Region, Russia\and
	INAF-Osservatorio Astronomico di Capodimonte, Salita Moiariello 16, 80131 Napoli, Italy \and
	School of Physics, University College Dublin, Belfield, Dublin 4, Ireland \and
	Max Planck Institute for Radio Astronomy, Auf dem H\"{u}gel 69, D-53121 Bonn, Germany}

\date{Received 16 February 2021 / Accepted 3 March 2023}

\abstract{The young stellar object (YSO) Th~28 possesses a highly collimated jet, which clearly exhibits an asymmetric brightness of its jet lobes at optical and near-infrared wavelengths. As with many other YSO outflows, there may be asymmetry in the physical parameters of the jet plasma in opposite jet lobes (e.g. electron density, temperature, and outflow velocity).
}
{We examined the Th~28 jet at high-spatial resolution in the regions where the jet material is collimated and accelerated. Our goal is to map the morphology and determine its physical parameters. We compared the results with those of other asymmetric YSO jets to determine the physical origin of such asymmetries.
}
{We used the integral field spectrograph SINFONI on the Very Large Telescope (VLT) of the European Southern Observatory to characterise the jet parameters in a $3\arcsec\times3\arcsec$ field around the central source of Th~28. We present high-resolution spectra of Th~28 covering the $JHK$ bands, obtained in June-July 2015.
}
{The images reveal gaseous structures out to distances of a few arcseconds around the stellar jet source. The [\ion{Fe}{II}] emission originates in highly collimated jet lobes. Two new axial knots are detected in the bipolar jet, one in each lobe, at angular distances of 1\arcsec{} in the blue lobe and 1\farcs2 in the red lobe. The H$_2$ radiation is emitted from an extended region with a radius of $\gtrsim270$ au, which is perpendicular to the jet. The position--velocity diagrams of the bright H$_2$ lines reveal faint H$_2$ emission along both jet lobes as well. The compact and faint HI emission (Pa$\beta$ and Br$\gamma$) comes from two regions, namely from a spherical region around the star and from the jet lobes. The maximum size of the jet launching region is derived as 0\farcs015, which corresponds to $\sim$3\,au at a distance of 185 pc, and the initial opening angle of the Th~28 jet is about 28\degr, which makes this jet substantially less collimated than most jets from other Classical T Tauri stars (CTTs).
}
{{The high-resolution SINFONI images show three groups of lines with different excitation conditions, which trace different gas structures. The emission in [\ion{Fe}{II}], H$_2$, and atomic hydrogen lines suggests a morphology in which the ionised gas in the disc (or at least very close to the jet launching site) appears to be disrupted by the jet. The resolved disc-like H$_2$ emission most likely arises in the disc atmosphere from shocks caused by a radial uncollimated wind. The asymmetry of the [\ion{Fe}{II}] photocentre shifts with respect to the jet source arises in the immediate vicinity of the driving source of Th~28 and suggests that the observed brightness asymmetry is intrinsic as well.
}}
\keywords{stars: jets, stars: variables: T Tauri, Herbig Ae/Be, stars: winds, outflows: individual: Th~28}

\titlerunning{SINFONI observations of the ThA~15-28 jet}
\maketitle
\section{Introduction}
\label{intro}
Many young stellar objects (YSOs) possess collimated outflows and/or jets, which are believed to allow the removal of excess angular momentum from their accretion discs \citep{Fer06,Pud07}. However, the complex picture of the transformation of circumstellar accretion into material ejection is not yet fully understood. Various theoretical models assume that the central jet engine should work in both directions in a similar way, producing a symmetrical bipolar jet \citep[e.g.][]{Fer04,Meli06,Pud07}. However, detailed studies have discovered that many of the bipolar YSO outflows exhibit remarkable asymmetries in the physical conditions seen in their opposite lobes on both large and small scales. Not only do some jets have lobes of different brightnesses, they also exhibit different physical parameters of the jet plasma in opposite lobes (e.g. electron density, temperature, radial velocity). For example, \citet{Hir94} found that 8 of the 15 bipolar Herbig-Haro (HH) outflows studied show velocity asymmetries between blue and redshifted outflows. Moreover, some objects reveal a single outflow component in only one direction, for example  the \object{LkH$\alpha$ 321} jet \citep{Cof04a} and \object{HH 34} \citep{Rei02}. 

\,The asymmetry can be intrinsic (i.e. caused by the `engine' itself) or extrinsic (i.e. caused by an inhomogeneous environment). Various mechanisms have been invoked to explain the observed jet asymmetry: variations in the morphology of the magnetic field near the central star \citep[e.g. in \object{HD 163296},][]{Was06}, warped circumstellar discs \citep[e.g. \object{HH 111},][]{Gom13}, and interactions between a stellar magnetic field and a circumstellar disc magnetic field \citep[e.g.][]{Mat12,Dyd15}. Recent numerical simulations of jets and winds from weakly ionised protoplanetary discs, taking into account non-ideal magnetohydrodynamic effects, also produce asymmetric ejections \citep[cf.][]{Bai17,Bet17}. The jet appearance is also affected by spatial extinction variations (i.e. extrinsic scenario), in particular shadowing the redshifted part by the disc--envelope. Admittedly, this does not cause differences in the physical parameters; proper extinction correction is provided.

The asymmetric properties of several young T~Tauri jets have been analysed using high-resolution observations in recent years. Using high-resolution spectra obtained with the Hubble Space Telescope, \citet{Mel09} found that the \object{RW Aur} jet has a strong asymmetry in electron density and temperature, excitation, and total hydrogen density between its opposite lobes. At the same time, the jet mass loss rate is similar in the two lobes, as in some other asymmetric jets, for example, \object{DG~Tau~B} \citep{Pod11} and HD~163296 \citep{Was06}. \citet{Mel09} concluded that the observed asymmetries of the RW Aur jet can be explained by different environmental conditions around the jet source and not by asymmetric jet formation. 

Using high-resolution Echelle spectra obtained with the HIRES spectrograph at the Keck~I telescope, \citet{Pod11} studied the asymmetry of the  \object{DG~Tau~B} jet properties, and concluded that it can be caused by a strong interaction of the jet lobes with an asymmetric ambient medium. \citet{Whi14} used the near-infrared (NIR) integral-field spectrograph on the Gemini telescope to study the structure of the asymmetric bipolar outflow of DG Tau. They concluded that the approaching jet is propagating in a low-density medium, while the receding jet is colliding with dense material along its path and appears as the large bubble-like structure visible in [\ion{Fe}{II}] 1.64 $\mu$m at 200 au from the source. \citet{Whi14} interpreted the properties of the observed H$_2$ emission as evidence for a clumpy medium above the circumstellar disc, which interacts with both the jet and the gaseous bubble. Finally, they concluded that the bipolar outflow from \object{DG Tau} is intrinsically symmetric, but the observed asymmetries are caused by an asymmetric distribution of the ambient circumstellar material. Three YSO jets have been studied during the last decade, RW Aur, DG Tau, and DG Tau B, which are visually asymmetric but seem to have intrinsic symmetry.

For this paper, we used the high resolution integral field spectrograph VLT/SINFONI to study another asymmetric bipolar microjet from the YSO \object{ThA 15-28} (\object{V1190 Sco}, \object{Sz 102}, hereafter Th~28) within 3\arcsec{} of the central source. Th~28 \citep[age $\sim$1-2 Myr,][]{Com10} is located in the \object{Lupus~3} cloud, at a distance of $\sim$185~pc\footnote{A much larger distance is suggested by the calibration of \citet{BJ18} based on Gaia DR2 \citep{Gaia18} parallax measurements, resulting in an estimated distance of $462_{-126}^{+278}$~pc.  However, because Th~28 is not a point source at optical wavelengths, this estimate may not be reliable. In Gaia DR3 \citep{Gaia20} the fidelity value for Th~28 is $\sim$0.05, whereas any parallaxes with a fidelity value of $\geqslant$0.5 are considered reliable. This confirms that the GAIA parallax for Th~28 is not trustworthy.} \citep{Gal13}.  The bipolar jet from Th~28, first discovered by \citet{Kra86}, exhibits a noticeable brightness asymmetry of its lobes, visible, for example, in [\ion{S}{II}] lines \citep{Wang09}. This collimated jet can be traced up to distances of several dozen arcseconds, with a few bow shocks visible along the jet axis on both sides of the source \citep{Wang09,Mur21}. The overall position angle (PA) of the flow is almost east--west, with the brighter red lobe of the microjet pointing towards the west. In a recent ALMA study, \citet{Lou16} mapped the CO lines in Th~28, and found a lower limit for the inclination of the disc rotation axis to the line of sight of $i > 73\degr$, and a jet inclination of $82\degr$. This confirmed the previous conclusion of \citet{Kra86} that the global inclination of the jet seems close to the plane of the sky ($i\sim82\degr$). \citet{Mur21} found that the inclination of the
two jet lobes is different: $i_{red} = 83\degr5$, whereas $i_{blue} = 77\degr$. Th~28 is also known as a source of X-ray emission \citep{Gon06}. Using the adaptive optics (AO) correction of the SINFONI integral field spectrograph, our objective was to obtain a spatial resolution $\sim$3 times higher than that achieved with the ISAAC slit spectrograph ($0\farcs5$) by \citet{Cof10}. In addition, the integral-field observations provide us with the opportunity to study the distribution of the emitting gas using spectroastrometric analysis.

The work is organised as follows. Details of SINFONI observations and data reduction are described in Sect.~\ref{obs}. In Sect.~\ref{res} we represent the main results of the analysis of the observations. In Sect.~\ref{em_33reg} we consider the general morphology of the Th~28 jet formation region, based on the analysis of different NIR (1.10--2.45 $\mu$m) spectral lines. In Sect.~\ref{jet_kin} we discuss the gas kinematics from an analysis of the [\ion{Fe}{II}] lines.  In Sect.~\ref{av_fe} we analyse the extinction A$_V$ calculated from the [\ion{Fe}{II}] line flux ratios.  In particular, in Sect.~\ref{xy_shift} we perform a spectroastrometric analysis and compute the photocentre shifts of the [\ion{Fe}{II}], Pa$\beta$, Br$\gamma$, and H$_2$ lines with respect to the central source.  In Sect.~\ref{h2_region} we analyse the morphology of the region that emits in the H$_2$ lines and present the physical parameters derived from the analysis of these molecular lines (Sect.~\ref{h2_param}). In Sect.~\ref{discuss} we report our spectral analysis of the jet and discuss our results. In Sect.~\ref{Conc} we summarise our findings and the scenario adopted for the jet.

\section{Observations and data analysis}
\label{obs}
\begin{figure}  
        \centering
        \resizebox{\hsize}{!}{\includegraphics{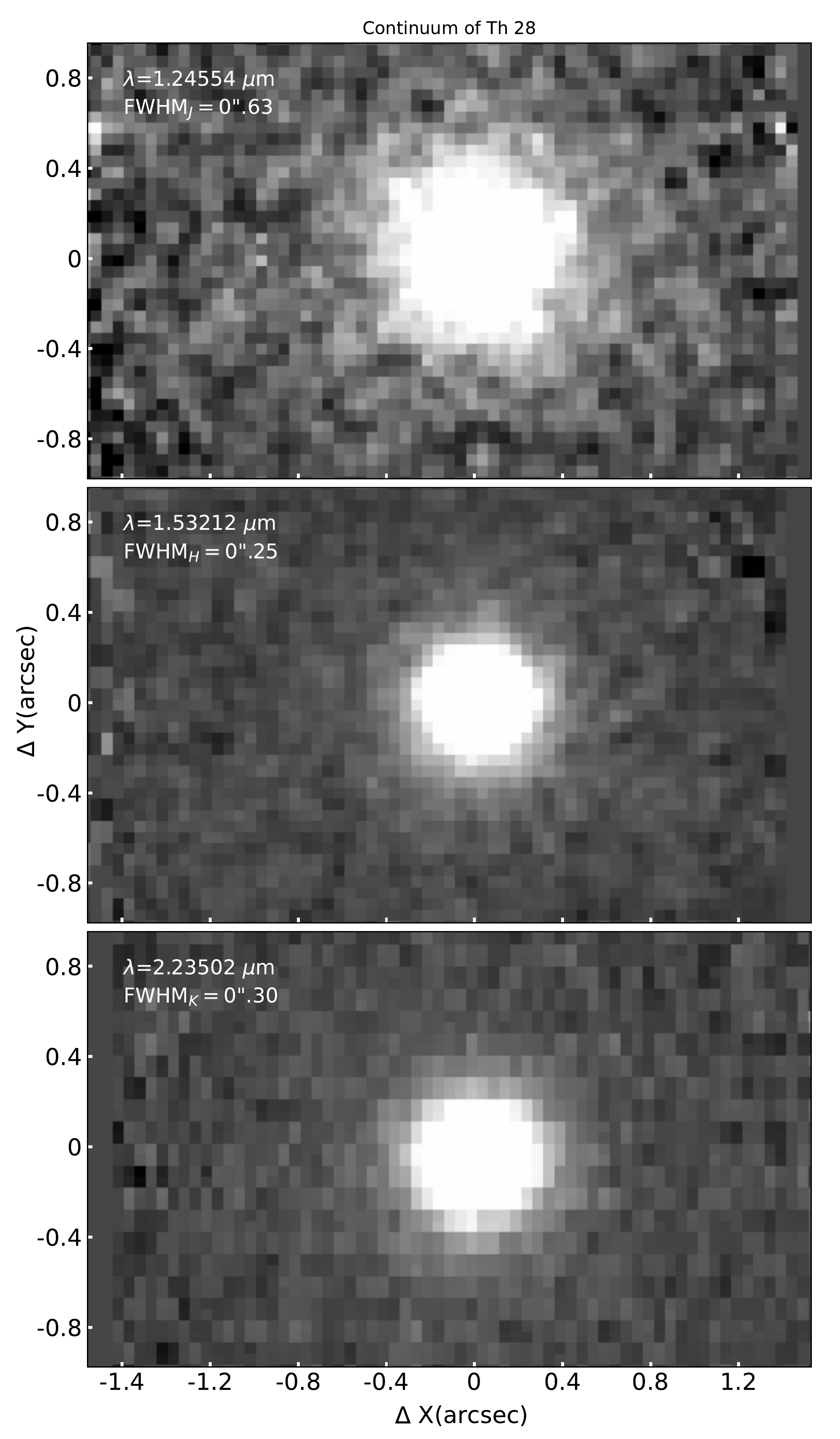}}
        \caption{Continuum images of Th~28 at the $J$, $H$, and $K$ band centres, with corresponding measured FWHMs of the continuum emission.}
        \label{cont}
\end{figure}

Th~28 (R.A.=16$^\mathrm{h}$ 08$^\mathrm{m}$ 30$^\mathrm{s}$ DEC=-39$\degr$ 03$\arcmin$ 11$\arcsec$) was observed with the SINFONI instrument at the VLT-UT4 telescope (ESO, Paranal, Chile) on June 4, 2015 ($J$, $K$), and July 17, 2015 ($H$). SINFONI is a NIR integral field spectrograph that works in combination with adaptive optics \citep{Eis03}; it was developed for observations of single faint objects. The Integral Field Unit (IFU) was used to obtain two-dimensional (2D) spectroscopy of the Th~28 jet in the wavelength bands $J$, $H$, and $K$, which provided spectral resolutions of $R=2\,000$ (150 km s$^{-1}$), 3\,000 (100 km s$^{-1}$), and 4\,000 (75 km s$^{-1}$), respectively. The wavelength ranges of the filters, excluding edge effects, are 1.10--1.40 $\mu$m for $J$, 1.45--1.85 $\mu$m for $H$, and 1.95--2.45 $\mu$m for $K$. The field of view (FoV) with adaptive optics (AO) mode is $3\arcsec\times3\arcsec$ with a spaxel size of $0\farcs05\times 0\farcs1$, where the detector direction corresponding to the larger pixel size was aligned with the jet axis of Th~28. The pipeline reduction rescales the cubes so that the final reduced cubes have pixels of the same size (0\farcs05) for both the $X$- and $Y$-directions. Since Th~28 is quite bright ($R = 15.4$ mag), it was used as a natural guide star for the AO system. A total integration time of 30~min per band was used. The seeing conditions during July 17 (DIMM=0\farcs6 for $H$ observations) were approximately twice better than during June 4 (DIMM$\simeq$1\farcs25 for $J$ and $K$ observations). The angular sizes of the central emission region in Th~28, derived from the continuum emission, amount to FWHM$_J \simeq$ 0\farcs63, FWHM$_H \simeq$ 0\farcs23, and FWHM$_K \simeq$ 0\farcs31 (single frames with continuum images are shown in Fig.~\ref{cont}).
The position angle (PA) of the SINFONI instrument was aligned with the jet PA (98\degr) so that the $Y$-direction in the observed frames coincides with the jet axis. To perform an absolute flux calibration, we observed the solar-type spectrophotometric standard HD~154504. Several telluric standards (HIP~080405, 082430, 090361) were also measured for atmospheric telluric correction.

The SINFONI integral-field spectroscopic observations produce raw science frames that carry both spatial and spectral information. Basic reduction of raw science images was performed according to the SINFONI standard data reduction recipes \citep{Mod07}, which include bad pixel removal, flat field, and dark correction. As a result, the frames were reconstructed as a 3D data cube, where every position in the $3\arcsec \times 3\arcsec$ field has a spectrum with a step of 2.5 \AA\ (0.00025 $\mu$m).

\begin{table*}
        \small\caption{\label{table_lines} Emission lines detected in SINFONI \textit{JHK} bands.}
        \setlength\tabcolsep{3pt} 
        \begin{tabular}{|l c r r | l c r r | l c r r |}
        \hline\hline\noalign{\smallskip}
        Element &  $\lambda_{{\rm vac}}$ & Flux$^*$ & Origin & Element &  $\lambda_{{\rm vac}}$ & Flux$^*$ & Origin & Element & $\lambda_{{\rm vac}}$ & Flux$^*$ & Origin \\
        & ($\mu$m) & $(\times10^{-16})$ & & & ($\mu$m) & $(\times10^{-16})$ & & & ($\mu$m) & $(\times10^{-16})$ & \\
    \hline\noalign{\smallskip}
    \multicolumn{4}{|c|}{$J$-band}                                        & [\ion{Fe}{II}]        & 1.5999 &  41.33\,$\pm$\,0.13 & jet  & \multicolumn{4}{|c|}{$K$-band}                                       \\
        \multicolumn{4}{|c|}{---------------------------------------------}    & Br$_{13}$      & 1.6119 &   1.65\,$\pm$\,0.03 &      & \multicolumn{4}{|c|}{---------------------------------------------}\\
        {[\ion{P}{II}]}    & 1.1886 &   6.97\,$\pm$\,0.04 &  disc?                    & Br$_{12}$      & 1.6416 &   3.24\,$\pm$\,0.04 &      & H$_2$ 1-0 S(3) & 1.9576 &103.66\,$\pm$\,0.29 & disc                \\
        {[\ion{Fe}{II}]}   & 1.2492 &   0.40\,$\pm$\,0.02 &                           & [\ion{Fe}{II}]        & 1.6440 & 243.30\,$\pm$\,0.41 & jet  & H$_2$ 1-0 S(2) & 2.0338 & 21.28\,$\pm$\,0.08 & disc                \\
        {[\ion{Fe}{II}]}   & 1.2570 & 296.58\,$\pm$\,0.29 &  jet                      & [\ion{Fe}{II}]        & 1.6642 &  20.28\,$\pm$\,0.08 & jet  & {[\ion{Fe}{II}]}      & 2.0468 &  5.04\,$\pm$\,0.03 &                     \\
        {[\ion{Fe}{II}]}   & 1.2708 &  16.72\,$\pm$\,0.07 &                           & [\ion{Fe}{II}]        & 1.6773 &  43.20\,$\pm$\,0.13 & jet  & He I           & 2.0605 &  1.47\,$\pm$\,0.01 &                     \\
        {[\ion{Fe}{II}]}   & 1.2791 &  37.36\,$\pm$\,0.08 &  jet?                     & Br$_{11}$      & 1.6817 &   7.83\,$\pm$\,0.08 &      & H$_2$ 1-0 S(1) & 2.1218 & 76.14\,$\pm$\,0.22 & disc                \\
        Pa$\beta$   & 1.2822 & 146.84\,$\pm$\,0.23 &disc+jet                   & H$_2$ 1-0 S(9) & 1.6877 &   2.48\,$\pm$\,0.03 &      & [\ion{Fe}{II}]        & 2.1338 &  1.73\,$\pm$\,0.02 &                     \\
        {[\ion{Fe}{II}]}   & 1.2946 &  48.55\,$\pm$\,0.09 &  jet?                     & [\ion{Fe}{II}]        & 1.7116 &   6.15\,$\pm$\,0.05 & jet  & Br$\gamma$     & 2.1661 &  8.57\,$\pm$\,0.14 & disc+jet?           \\
        {[\ion{Fe}{II}]}   & 1.2985 &   9.15\,$\pm$\,0.04 &                           & H$_2$ 1-0 S(8) & 1.7147 &   0.33\,$\pm$\,0.01 &      & H$_2$ 1-0 S(0) & 2.2233 & 12.66\,$\pm$\,0.05 & disc                \\
        {[\ion{Fe}{II}]}   & 1.3209 &  72.29\,$\pm$\,0.10 &  jet                      & Br$_{10}$      & 1.7371 &   6.39\,$\pm$\,0.07 &      & [\ion{Fe}{II}]        & 2.2244 &  7.51\,$\pm$\,0.04 & jet                 \\ 
        {[\ion{Fe}{II}]}   & 1.3284 &  21.15\,$\pm$\,0.06 &                           & [\ion{Fe}{II}]        & 1.7454 &   8.98\,$\pm$\,0.06 & jet  & H$_2$ 2-1 S(1) & 2.2477 &  0.71\,$\pm$\,0.01 &                     \\
        \multicolumn{4}{|c|}{$H$-band}                                         & H$_2$ 1-0 S(7) & 1.7480 &   7.71\,$\pm$\,0.04 & disc & H$_2$ 1-0 Q(1) & 2.4066 & 89.31\,$\pm$\,0.26 & disc                \\
        \multicolumn{4}{|c|}{---------------------------------------------}    & [\ion{Fe}{II}]        & 1.7489 &   1.13\,$\pm$\,0.01 & jet  & H$_2$ 1-0 Q(2) & 2.4134 &  9.70\,$\pm$\,0.04 & disc                \\
        {[\ion{Fe}{II}]}   & 1.5339 & 60.74\,$\pm$\,0.20  & jet                       & H$_2$ 1-0 S(6) & 1.7879 &   1.36\,$\pm$\,0.06 &      & H$_2$ 1-0 Q(3) & 2.4237 & 56.65\,$\pm$\,0.18 & disc                \\ 
        Br$_{17}$   & 1.5443 &  0.31\,$\pm$\,0.15  &                           & [\ion{Fe}{II}]        & 1.7976 &  16.30\,$\pm$\,0.08 & jet  & H$_2$ 1-0 Q(4) & 2.4375 &  3.74\,$\pm$\,0.08 & disc                \\
        Br$_{16}$   & 1.5562 &  0.81\,$\pm$\,0.02  &                           & [\ion{Fe}{II}]        & 1.8005 &  17.29\,$\pm$\,0.10 & jet  &                &        &                    &                     \\
        Br$_{14}$   & 1.5885 &  1.52\,$\pm$\,0.03  &                           & [\ion{Fe}{II}]        & 1.8099 &  35.50\,$\pm$\,0.12 & jet  &                &        &                    &                     \\
        \hline\noalign{\smallskip}
        \end{tabular}
        
        \textbf{Notes.} $^{(*)}$ The flux is integrated over the continuum-subtracted spectrum and in erg s$^{-1}$ cm$^{-2}$ units. For the H$_2$ lines the flux is summed for the bright central core with a size of $\sim$1\farcs5 and does not include the faint periphery. For [\ion{Fe}{II}] lines the bright areas of emission with S/N$ > 3$ are integrated along the jet lobes.
\end{table*}                                  

The wavelength calibration based on observations of a xenon--argon arc lamp led to high systematic errors. Therefore, for this purpose we used telluric lines, which provide a better calibration accuracy \citep[e.g.][]{Agra14}. The estimated uncertainties of the measured wavelengths after using the telluric lines are $\sim 0.1-0.3$\,\AA\ (ranging between $\pm 1-10$ km s$^{-1}$, depending on the wavelength). The estimated FWHM of the telluric lines at $J$, $H$, and $K$ are $205\pm33$, $155\pm16$, and $139\pm30$ km s$^{-1}$, respectively. All radial velocities are systemic and measured with respect to the Local Standard of Rest (LSR) ($V_\mathrm{LSR}$ = 3 km~s$^{-1}$), which was obtained from recent ALMA observations by \citet{Lou16}. Absolute flux calibration was performed by measuring the density flux (in counts) on the standard star by fitting a 2D Gaussian profile at each wavelength. The detector count rates were then converted to erg s$^{-1}$ cm$^{-2}$ $\mu$m$^{-1}$ by comparison with the reference spectrum of the spectral standard. Telluric standard stars were used to correct the spectrum for the detector spectral response. Standard--telluric stars and the science target were observed at similar airmasses, so no atmospheric extinction corrections were applied.  Finally, we subtracted the continuum emission (a combination of stellar, dust, and scattered emission) at each position in the field by fitting a third-order polynomial to the continuum at each point.

\section{Results}
\label{res}
\begin{figure*} 
        \centering
        \resizebox{\hsize}{!}{\includegraphics{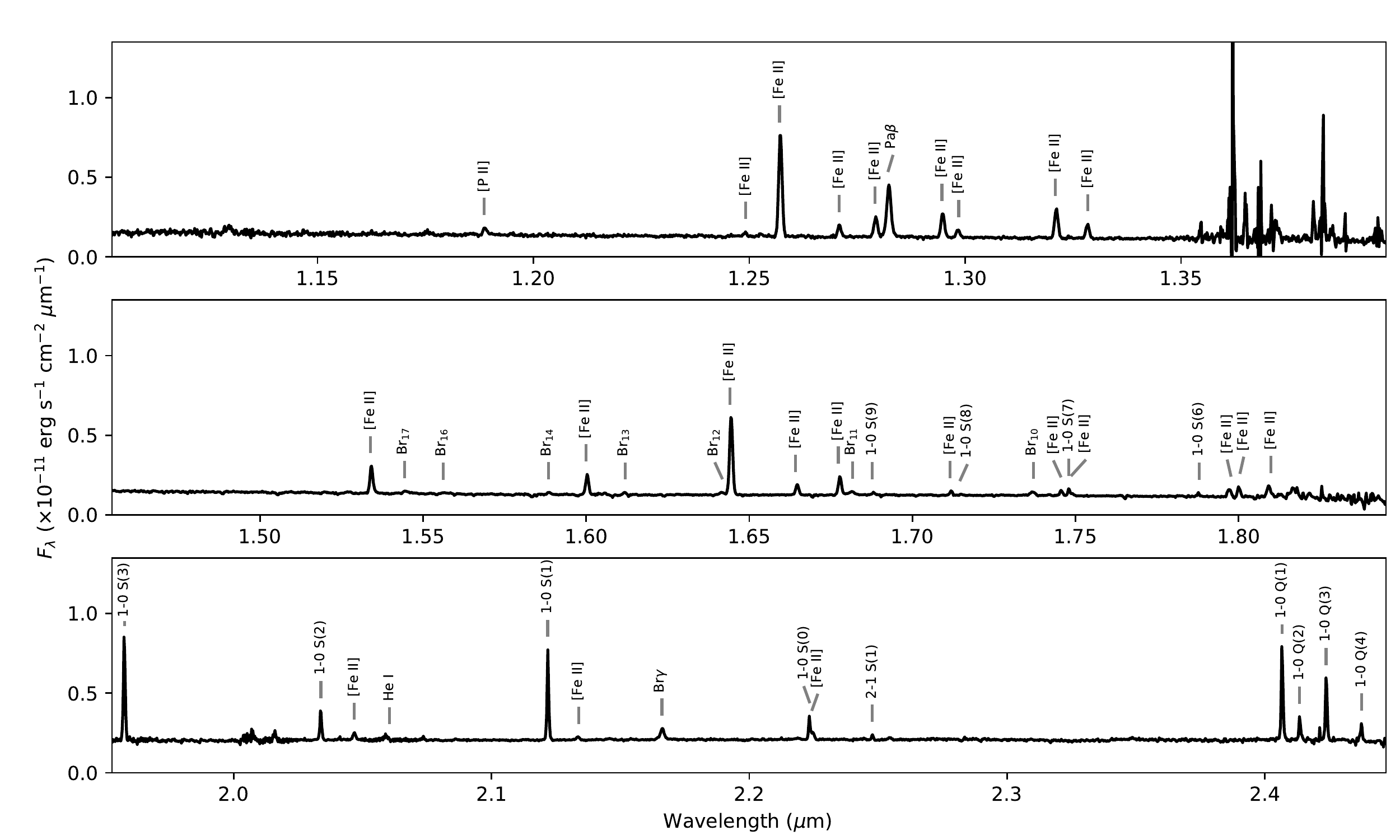}}
        \caption{1D spectra in the $J$, $H$, and $K$ bands, summed over the central $0\farcs5 \times 0\farcs5$ of Th~28.}
        \label{em_lines}
\end{figure*}

We inspected the data cubes and identified all emission features found in the $J$, $H$, and $K$ bands (see Table~\ref{table_lines}). The average 1D spectra in the central $0\farcs5 \times 0\farcs5$ region are shown in Fig.~\ref{em_lines}, where the identified emission lines are marked. These emission lines represent three main groups: forbidden [\ion{Fe}{II}] lines, molecular hydrogen lines (H$_2$), and atomic hydrogen lines. Atomic H lines are mostly represented by Brackett-series recombination lines. The $K$ band contains faint Br$\gamma$ emission, and the lines from Br$_{10}$ to Br$_{16}$ are also detected in the $H$ band; Br$\delta$ and Br$_9$ are not detected as they fall into the wavelength gap between the $H$ and $K$ bands. In the $J$ band, the Pa$\beta$ line is detected. The brightest H$_2$ lines are found in the $K$ band, but some molecular hydrogen emission is also detected in the $H$ band. Two emission lines that do not belong to these large groups were identified as [\ion{P}{II}] 1.1886 $\mu$m and \ion{He}{I} 2.0605 $\mu$m, which were also detected in the spectrum of the YSO jet \object{HH 99B} \citep{Gia08}. The vacuum wavelengths of the identified lines are listed in Table~\ref{table_lines}, and the origin of the line is identified based on their morphology.

\subsection{Jet morphology in atomic lines in the $3\arcsec\times 3\arcsec$ region}
\label{em_33reg}

\begin{figure*} 
        \centering
        \resizebox{\hsize}{!}{\includegraphics{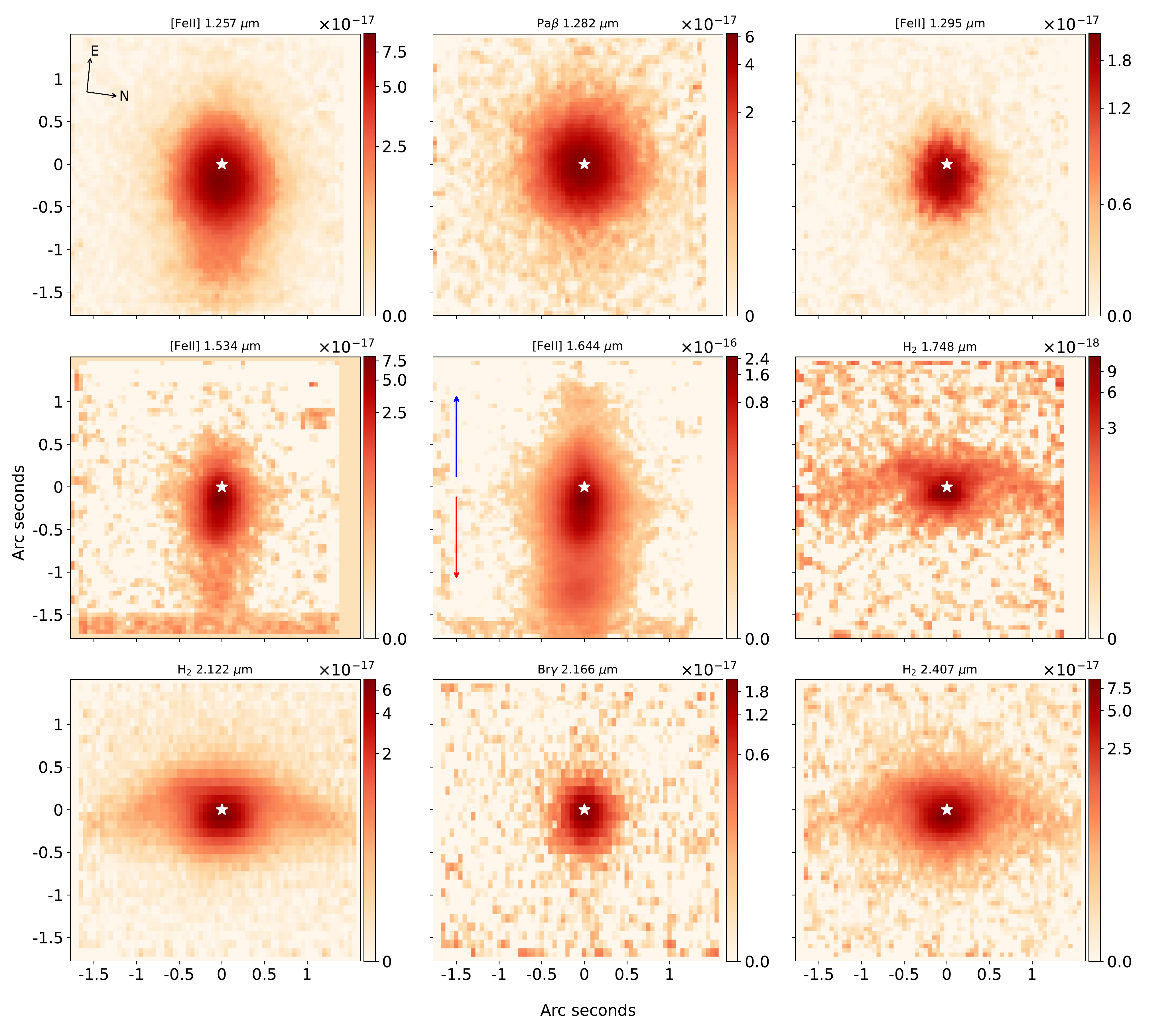}}
        \caption{Total line flux maps of the nine brightest emission lines in Th~28 over the $J$, $H$, and $K$ bands. Each map shows the total line flux integrated over all the velocity frames where emission is detected. The star  gives the position of the central source, calculated from the continuum (stellar) profile. The coloured arrows indicate the direction of the blue and red lobes. The vertical log-scale shows the fluxes in erg s$^{-1}$ cm$^{-2}$.}
        \label{lines9}
\end{figure*}

\begin{figure*}
        \centering
        \resizebox{\hsize}{!}{\includegraphics{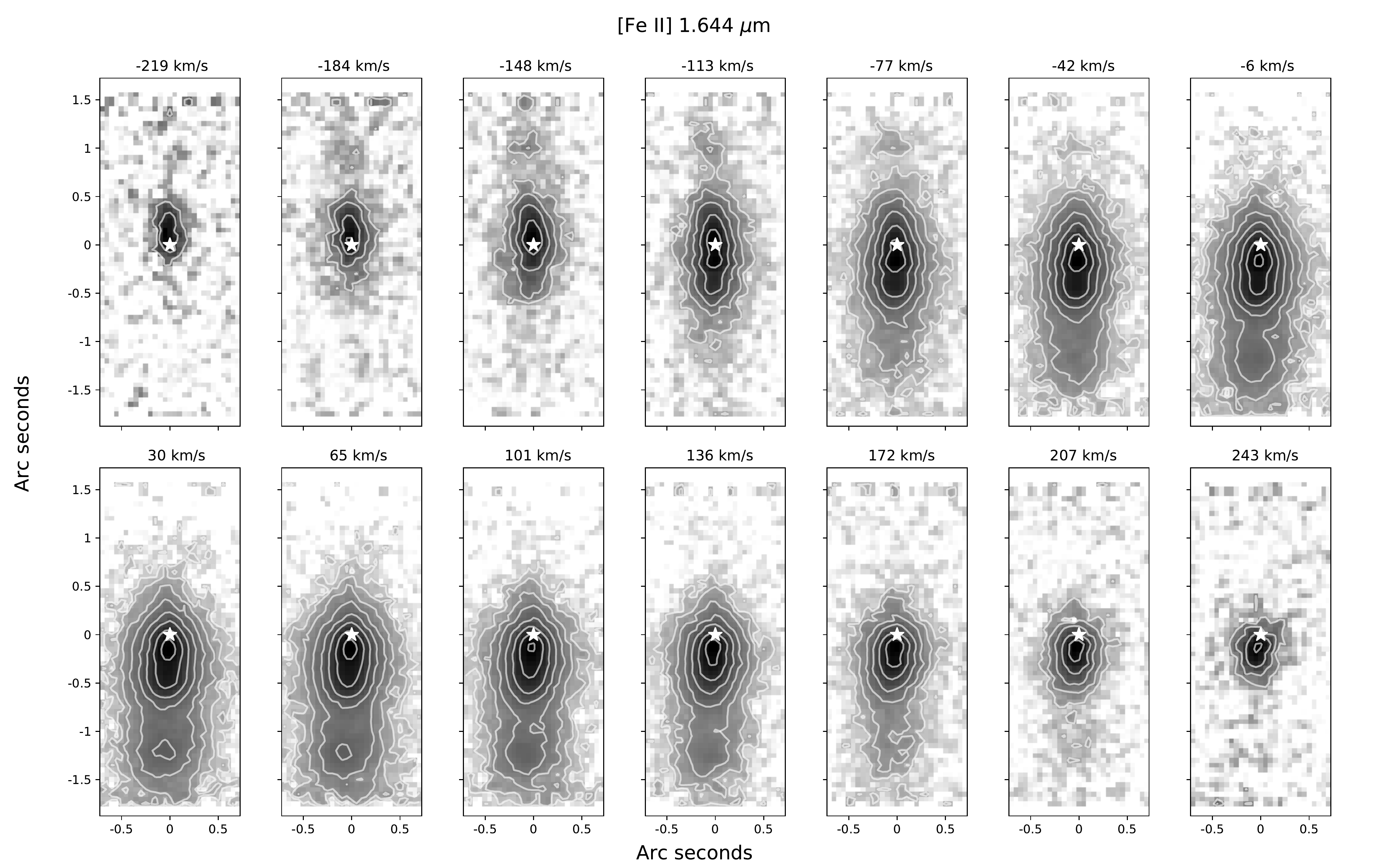}}
        \caption{Continuum-subtracted channel maps for [\ion{Fe}{II}] 1.644 $\mu$m. The 14 panels show that the [\ion{Fe}{II}] emission covers a wavelength range from $-220$ km~s$^{-1}$ to 243 km~s$^{-1}$. The peak position of the continuum flux is indicated  by a white star. Contours in units of erg s$^{-1}$ cm$^{-2}$ $\mu$m$^{-1}$ consist of nine  levels from $2\times10^{-15}$ to $3.2\times10^{-13}$.}
        \label{Feii_14}
\end{figure*}

Figure~\ref{lines9} shows the total flux maps of the brightest emission lines, which trace different gaseous structures in the vicinity of Th~28.  First, we note that the bright [\ion{Fe}{II}] emission ([\ion{Fe}{II}] 1.257, 1.534, 1.644~$\mu$m) seems to originate in the jet lobes, since its elongation coincides with the jet axis. With the PA=98\degr\ used for SINFONI, the upper jet lobe in Fig.~\ref{lines9} corresponds to the blueshifted (east)  lobe, while the lower lobe (i.e. the brightest jet component) corresponds to the redshifted component (west). The maps also indicate that the [\ion{Fe}{II}] emission has the same asymmetric shape between the red and blue lobes as seen in the [\ion{S}{II}] lines \citep{Cof10,Wang09}. However, the fainter [\ion{Fe}{II}] 1.295~$\mu$m ($J$ band) is rounder and is reminiscent of the shape of Pa$\beta$. At the same time, although the shape of the bright [\ion{Fe}{II}] 1.257~$\mu$m emission is wider than the [\ion{Fe}{II}] lines in the $H$ band (obtained with better seeing), it still traces the shape of the jet. Therefore, the [\ion{Fe}{II}] 1.295~$\mu$m line seems also to originate in the jet, but its emission is bright enough for detection only in the immediate vicinity of the jet source.  The bright [\ion{Fe}{II}] lines show two knots at similar distances in opposite directions. Almost all bright iron lines in the redshifted lobe show an emission knot at 1\farcs2 from the central source, while the brightest [\ion{Fe}{II}] 1.644 $\mu$m line reveals a faint knot at 1\arcsec{} in the blueshifted beam (Fig.~\ref{lines9} and Fig.~\ref{Feii_14}).

Unlike the [\ion{Fe}{II}] lines, both the Pa$\beta$ and Br$\gamma$ emitting regions are mostly round and symmetric, with brightness maxima roughly coincident with the position of the stellar continuum. Measurement of the continuum-subtracted emission region shows that the Pa$\beta$ line is much brighter and more extended than the Br$\gamma$ line. We performed Moffat fitting of the 1D profiles along the jet axis ($Y$-direction) and perpendicularly to this direction ($X$-direction) to determine if there is any departure of the Pa$\beta$- and Br$\gamma$-emitting regions from a circular shape. The fit of the Pa$\beta$ line intensity distribution shows that it is generally somewhat round (FWHM$_X$ = 0\farcs66$\pm$0\farcs03, FWHM$_Y$ = 0\farcs70$\pm$0\farcs04), while the angular size of the Br$\gamma$ emitting region is FWHM$_X$ = FWHM$_Y$ = 0\farcs33$\pm$0\farcs01.

We compared these values with the FWHM of the continuum profiles to check whether the \ion{H}{I} emitting regions are spatially resolved. For this goal, we computed the averaged continuum FWHM from 15 single frames from both sides of both lines (in total, 30 frames for each line). The distribution of these FWHM values is flat and does not show any wavelength dependence for these spectral regions. We performed a Moffat 1D fitting of the stellar profile of the continuum and found that for the Pa$\beta$ region, the continuum has a similar angular size (i.e. FWHM$_X$ = 0\farcs67$\pm$0\farcs05 and FWHM$_Y$ = 0\farcs69$\pm$0\farcs05). For the Br$\gamma$ region, the fitted continuum profile had FWHM$_X$ = 0\farcs33$\pm$0\farcs01 and FWHM$_Y$ = 0\farcs32$\pm$0\farcs01. Therefore, the results imply that both the Pa$\beta$- and Br$\gamma$-emitting regions are not resolved. This agrees with the results of \citet{Car16} for the massive YSO \object{IRAS~13481-6124} (also with a jet), where the Br$\gamma$-emitting region is very compact. However, \citet{Car16} used interferometric techniques and concluded that the Br$\gamma$ region is extended, tracing a jet-like structure.

At the same time, Fig.~\ref{lines9} shows very faint emission along the jet lobes in both Br$\gamma$ and Pa$\beta$. The position--velocity (PV) diagram of the Pa$\beta$ line from \citet{Cof10} also confirms that some part of this emission can come from the jet beam, primarily from the redshifted lobe. We calculated a PV diagram of Pa$\beta$ for the SINFONI data and also found that both jet lobes show faint emission in this line. Our results also indicate that the red Pa$\beta$ lobe is brighter than the blue one, which agrees with \citet{Cof10}. The 1D fitting of the Pa$\beta$ and Br$\gamma$ integral spectra shows that the two lines have similar and wide profiles, with FWHM$_{in}$(Br$\gamma$)=225~km~s$^{-1}$ and FWHM$_{in}$(Pa$\beta$)=220~km~s$^{-1}$. Both lines are redshifted with respect to the system velocity, but the peak of the Br$\gamma$ emission is at $+18\pm 4$ km~s$^{-1}$, whereas the peak of the Pa$\beta$ emission is at $+34\pm 3$ km~s$^{-1}$. These results agree with the peak radial velocities (RVs) derived by \citet{Cof10} for the slit parallel to the jet axis for Pa$\beta$ and Br$\gamma$ in the red lobe. 

\begin{figure}
        \centering
        \resizebox{\hsize}{!}{\includegraphics{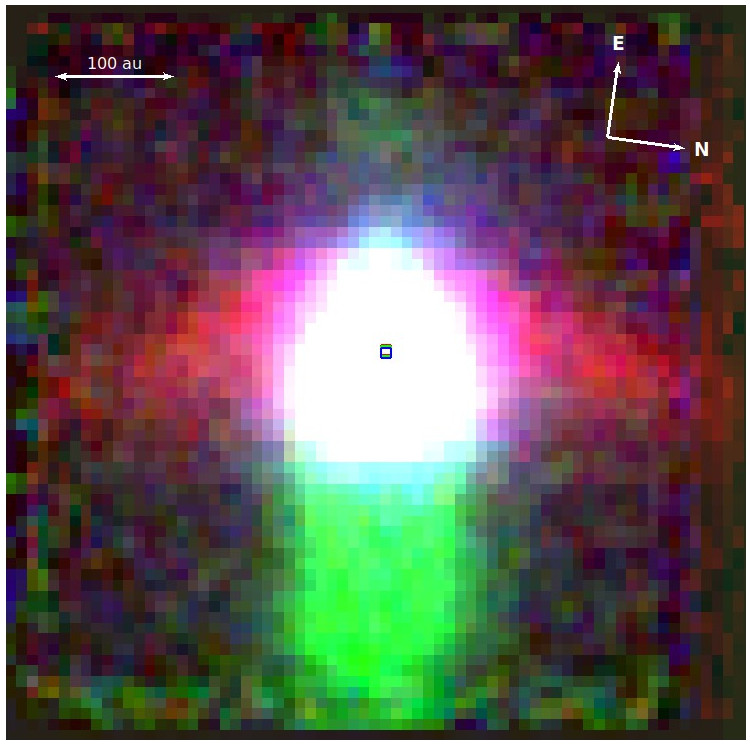}}
        \caption{Tricolour map of the $3\arcsec \times 3\arcsec$ field around Th~28, combined from three emission lines outlining different emission regions: \textit{blue} (Pa$\beta$), \textit{green} ([\ion{Fe}{II}] 1.644 $\mu$m), and \textit{red} (H$_2$ 2.122 $\mu$m). The plane of H$_2$-emitting gas is clearly orthogonal to the position of jet axis outlined by forbidden [\ion{Fe}{II}]-emission. Three small squares give the position of the star calculated from the continuum stellar profile.}
        \label{lines3}
\end{figure}

We show the composite image of Th~28 of three bright emission lines, Pa$\beta$, [\ion{Fe}{II}] 1.644~$\mu$m, and H$_2$ 2.122~$\mu$m, in Fig.~\ref{lines3}. The fluxes of all the lines were integrated over all velocity bins with line emission. Three differently coloured small squares give the position of the continuum stellar profile near each line and perfectly match within the uncertainties. As can be seen, the plane of the disc-like emitting H$_2$ area is orthogonal to the jet axis. The faint emission knot in [\ion{Fe}{II}] in the upper (blueshifted) jet lobe is also visible in the combined image.

\subsection{Jet kinematics}
\label{jet_kin}

Figure~\ref{Feii_14} shows the velocity channel maps, with the radial velocity distribution of continuum-subtracted [\ion{Fe}{II}] emission from the 1.644 $\mu$m line. The 14 frames clearly exhibit a complex behaviour of gas emission in both the red- and blueshifted jet lobes. First, the gas has an asymmetrical velocity distribution. The RV extends up to $-220$ km~s$^{-1}$ in the blueshifted lobe, while in the redshifted lobe it is more extended and reaches up to 243 km~s$^{-1}$. Interestingly, the redshifted jet lobe becomes visible at negative RVs (i.e. the bipolar jet looks nearly symmetrical at RV = $-113$ km~s$^{-1}$). The same effect can also be seen in the opposite direction: emission from the blueshifted lobe is still detectable at redshifted velocities (within $0\farcs5$ of the central source). Since the jet is located near the plane of the sky, this  behaviour implies that we register the reverse velocity components of the emission, which can form, for instance, from divergent flows if the jet opening angle is larger than the inclination of the jet to the sky plane. Another source of the reversal components could be the possible contribution from scattering radiation by the gas--dust structures of the jet. The FWHM of the instrumental spectral profile in the $H$ band is $\sim$110 km s$^{-1}$ and is spread over three to four channel maps, spaced by 36 km s$^{-1}$. This also influences the appearance of the radiation from the outer line wings. The emission maps also show a displacement of the photocentre with respect to the stellar position, which is discussed in Sect.~\ref{xy_shift}.
\begin{figure*}
\centering
\begin{minipage}[c]{3.1cm}
        \resizebox{\hsize}{!}{\includegraphics{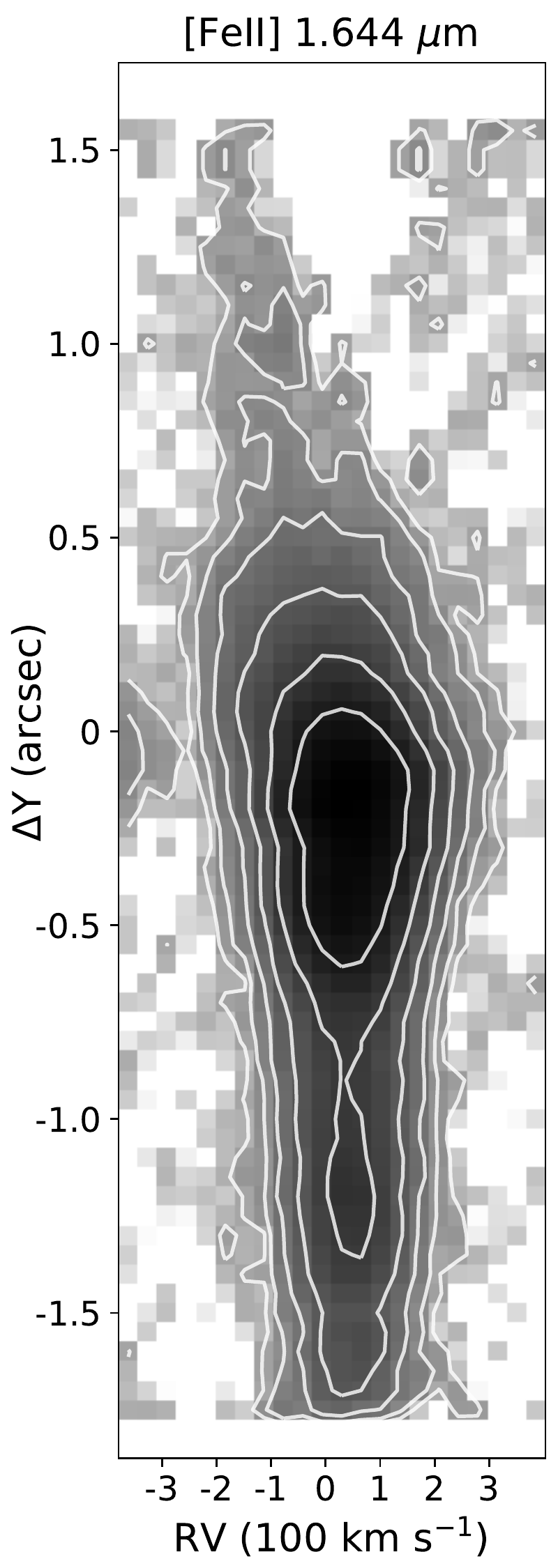}}
\end{minipage}
\begin{minipage}[c]{3.1cm}
        \resizebox{\hsize}{!}{\includegraphics{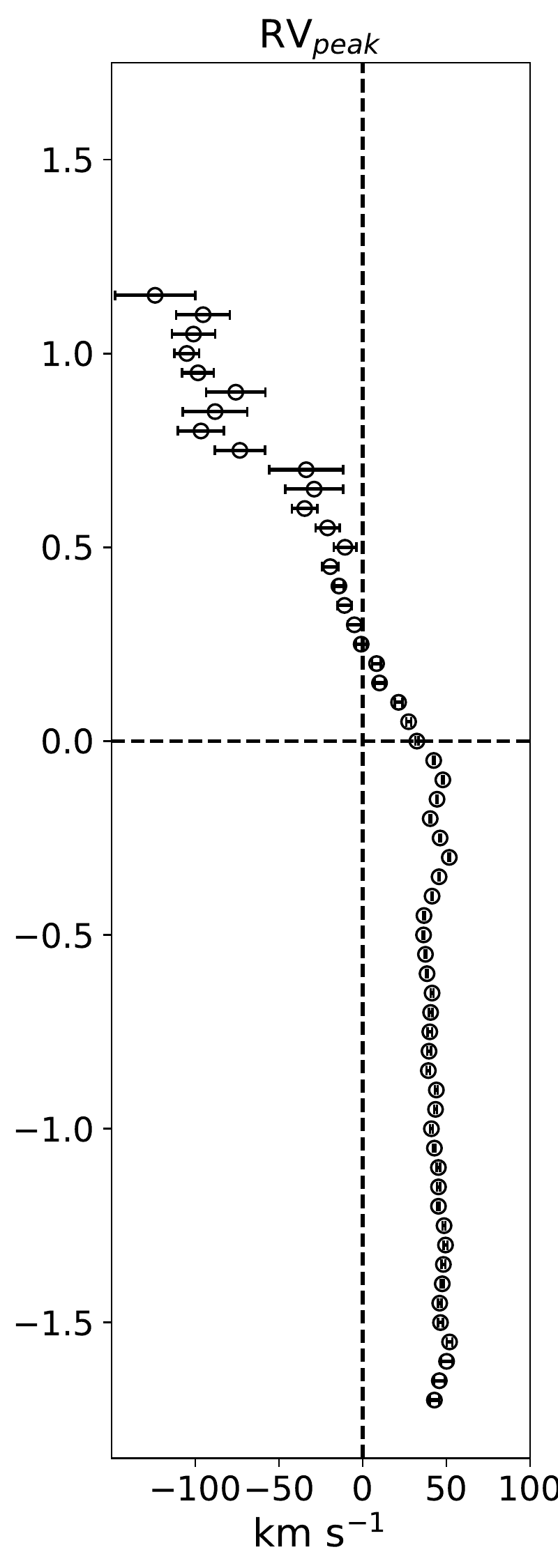}}
\end{minipage}
\begin{minipage}[c]{3.1cm}
        \resizebox{\hsize}{!}{\includegraphics{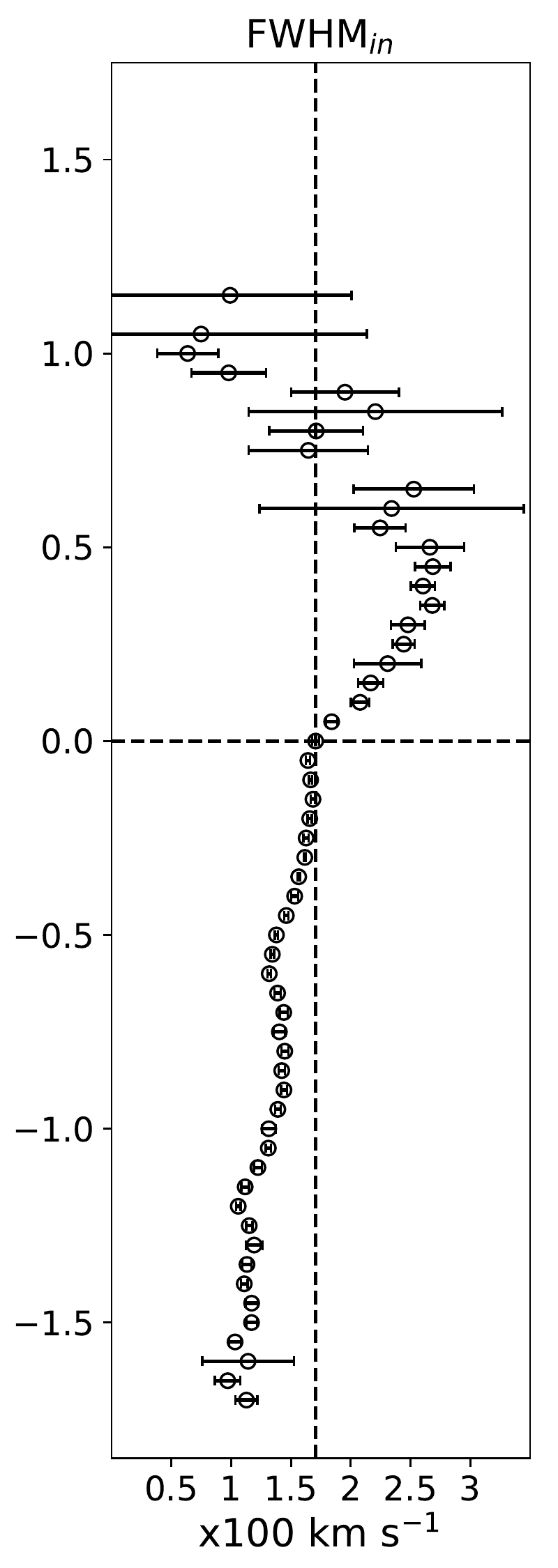}}
\end{minipage}
\begin{minipage}[c]{8.6cm}
        \resizebox{\hsize}{!}{\includegraphics{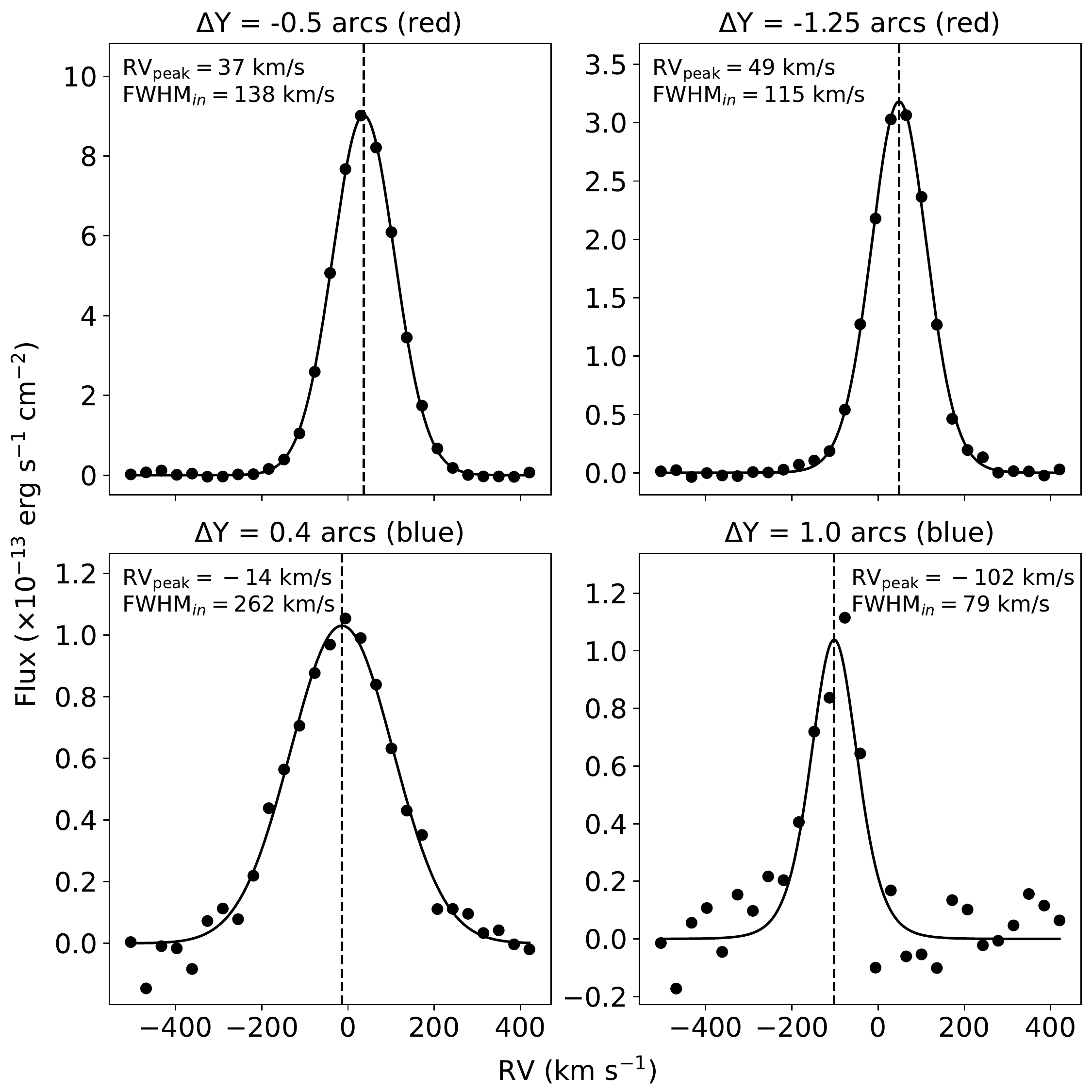}}
\end{minipage}
\caption{Distribution of the [Fe II] 1.644 $\mu$m emission along the jet. \textit{Left image}: PV diagrams of [\ion{Fe}{II}] 1.644 $\mu$m line in the Th~28 jet, obtained by coadding fluxes in the 16 spectral frames across the jet. Logarithmic contours are drawn from $10^{-14}$ erg s$^{-1}$ cm$^{-2}$ arcsec$^{-2}$ \AA$^{-1}$, with an incremental logarithmic step of 0.3 dex. \textit{Two central panels}:  RV$_{peak}$ and FWHM$_{in}$ derived from fitting the PV diagram as a function of the distance from the source. The vertical line in the FWHM panel corresponds to 170 km s$^{-1}$ measured for the position of the source. \textit{Right four panels}: Spectral profiles across the jet at selected position in the blue and red lobes (\textit{black points}). The data was fit using a Moffat function (\textit{solid line}). The intrinsic FWHM and the peak RV are shown at the top of each panel.}
        \label{pv_feii}
\end{figure*}

To calculate the maximum RVs of the jet material, we constructed a PV diagram for the [\ion{Fe}{II}] 1.644 $\mu$m line (Fig.~\ref{pv_feii}). The distributions of RV$_{peak}$ and FWHM$_{in}$ derived from fitting the PV diagram are shown as a function of distance from the source. The figure also shows the spectral profiles derived at selected positions and a fit of the line profiles using a Moffat function. The fitting shows that the kinematics of the [\ion{Fe}{II}] emission along the entire redshifted lobe is quite stable, with a peak velocity of $+44\pm7.7$ km s$^{-1}$ at a distance of 1\farcs7. However, the blue jet has a different morphology and the peak blueshifted velocities are not as evident or as stable as those of the redshifted velocities. Our fitting shows that the RVs gradually increase from low values at the positions closest to the central source to high absolute values at far distances, even exceeding the velocities derived in the receding lobe. For positions up to +0\farcs5, the peak RVs do not exceed $-20$ km s$^{-1}$. For example, RV$_{peak} = -14$ km s$^{-1}$ at $\Delta$Y = +0\farcs4 (Fig.~\ref{pv_feii}), but at greater distances ($>+0\farcs5$) the gas in the approaching lobe reaches large negative RVs. The panel for $\delta= +1\farcs0$ in Fig.~\ref{pv_feii} shows the emission line profile in the faint blueshifted knot, and was summed over 3 pixels, with an average distance from the central source of 1\farcs0. The peak RV in the knot is $\sim -100$ km s$^{-1}$ (i.e. twice as large as in the opposite direction). For the blue lobe, the derived FWHM$_{in}$ varies from large values ($\sim$270 km s$^{-1}$) close to the source to $\sim$ 80 km s$^{-1}$ at the most distant measured positions. This also differs from the FWHM distribution in the red lobe, where the FWHMs of [\ion{Fe}{II}] 1.644 $\mu$m are similar and range between 90 and 150 km s$^{-1}$ for all positions along the red lobe. The FWHM$_{in}$ measured at the source position corresponds to 170 km s$^{-1}$.

\subsection{Determination of $A_V$ from [\ion{Fe}{II}] lines}
\label{av_fe}

The [\ion{Fe}{II}] lines from the same upper excitation state are often used to determine the visual extinction $A_V$. Their intrinsic ratios are determined by the Einstein $A$ coefficient (which corresponds to the case of $A_V$ = 0) and are not affected by the physical conditions in the gas plasma. We used the [\ion{Fe}{II}] 1.644, 1.321, and 1.257~$\mu$m lines to determine $A_V$ as they all form with an upper level of $\mathrm{a}^4\mathrm{D}_{7/2}$ and are bright in the Th~28 spectra. The intrinsic [\ion{Fe}{II}] 1.644/1.257 and 1.644/1.321 ratios, derived from the Einstein $A$ coefficient, correspond to 0.88 and 3.13 \citep{SH06}, respectively. To calculate $A_V$ from the line ratio, we used the extinction law of \citet{ccm89}.

\begin{figure}
        \centering
        \resizebox{\hsize}{!}{\includegraphics{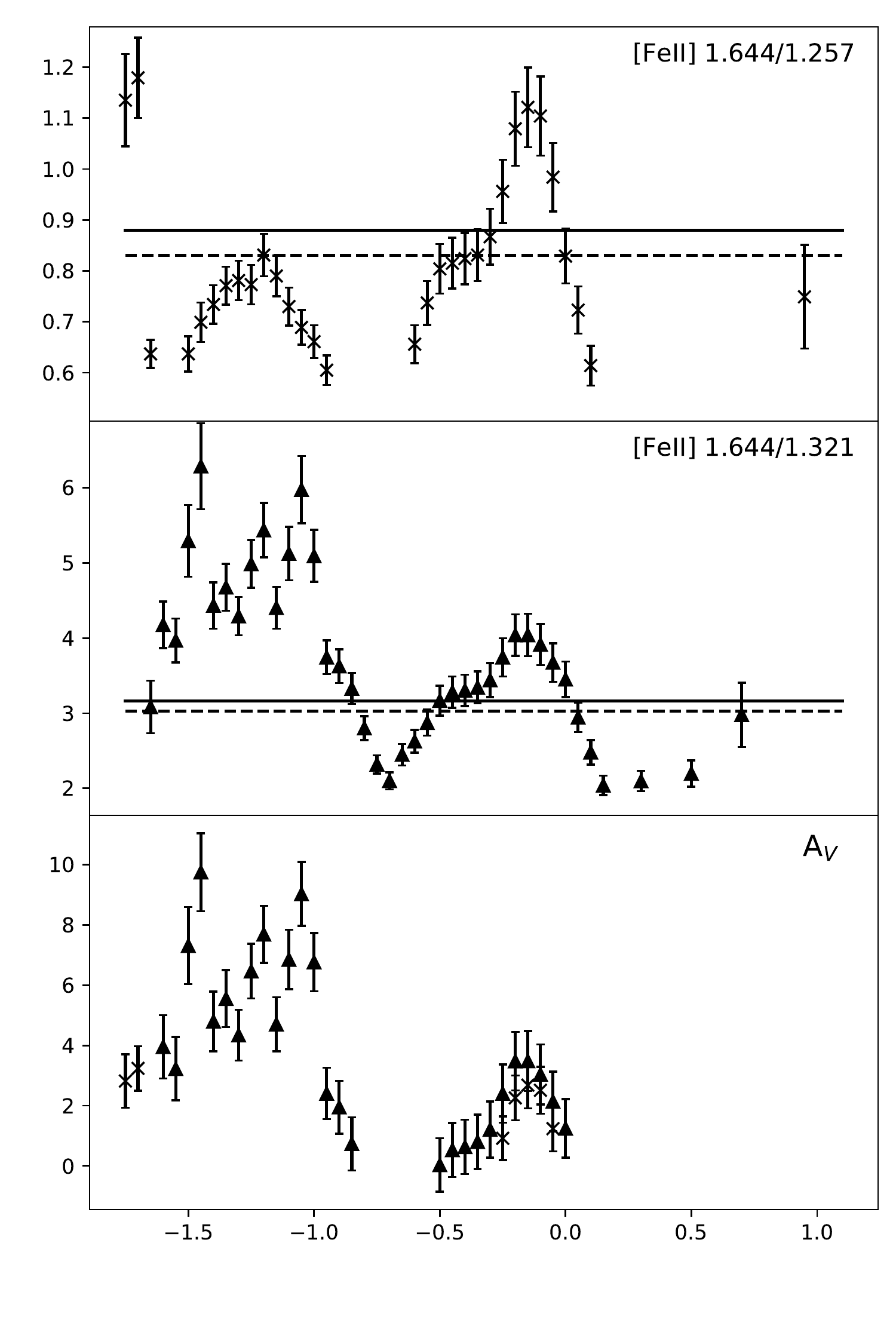}}
        \caption{Visual extinction $A_V$ along the Th~28 jet calculated from [\ion{Fe}{II}] 1.644/1.257 and 1.644/1.321. The theoretical intrinsic ratio for $A_V=0$ from \citet{SH06} is shown by the \textit{solid line}, whereas the empirical intrinsic ratio determined by \citet{Gia15} is shown as the dashed line. The error bars are calculated from the RMS errors plus an uncertainty of  30\% for the determination of the Einstein $A$ coefficient and an uncertainty of 10\% linked with the flux calibration. The extinction $A_V$ (\textit{lower panel}) is calculated for all positions, showing the ratio value above the theoretical intrinsic [\ion{Fe}{II}] ratio.}
        \label{ratio_av}
\end{figure}

In our case, however, the [\ion{Fe}{II}] 1.644 and 1.257~$\mu$m lines were obtained under different observation conditions, and the observed ratio appears to be affected by different seeing. To eliminate problems due to different seeing and PSF, we integrated the flux of these [\ion{Fe}{II}] lines across the width of the jet and calculated the [\ion{Fe}{II}] 1.644/1.257 ratio along the jet (Fig.~\ref{ratio_av}). The obtained ratios (within the estimated errors) were mostly lower than the intrinsic ratio of \citet{SH06}, and there are a few positions near the jet source in the red lobe where this line ratio is higher than the intrinsic one, $1.05\pm0.05$ which corresponds to $A_V \approx 2$. We also note that the [\ion{Fe}{II}] 1.644/1.257 value averaged over the entire jet beam is $0.72\pm 0.02$, which is consistent within the errors with that of the P Cygni nebula for $A_V$ = 0 calculated using the Einstein $A$ coefficient from \citet{NS88} \citep{SH06}. Unfortunately, there is a large uncertainty in determining the Einstein $A$ coefficient, which reaches more than 30\% by computing them with different methods \citep{Gia08}. Recently, \citet{Gia15} performed an empirical determination of the intrinsic ratios of several strong [\ion{Fe}{II}] lines in the spectrum of the bright HH object \object{HH~1}. They found that the ratios agree better with the predictions of the relativistic Hartree-Fock model of \citet{Qui96}. However, our observed [\ion{Fe}{II}] 1.644/1.257 ratio is lower than those determined for HH~1 for the $A_V$ = 0 case at many of these positions along the Th~28 jet.

To calculate the [\ion{Fe}{II}] 1.644/1.321 ratio (Fig.~\ref{ratio_av}), we again used the [\ion{Fe}{II}] fluxes integrated perpendicularly across the jet and calculated the ratio along the jet lobes. The ratio values are above the intrinsic ratios in the red lobe at the two sections, between -0.5 and 0 arcsec and between -1.6 and -0.85. Although different authors \citep{Nis05,Pod06} point out that the [\ion{Fe}{II}] 1.644/1.257 ratio overestimates A$_V$ compared to that calculated with [\ion{Fe}{II}] 1.644/1.317, the two ratios give almost the same A$_V \approx 2^m$ between -0.5 and 0 arcsec (the red lobe). However, [\ion{Fe}{II}] 1.644/1.317 between -1.6 and -0.85 gives much higher ratios that allow us to calculate A$_V=5^m-6^m$, which is higher than that measured at -0.5--0 arcs. This is unusual, however, some scattering of the [Fe II] emission in the central regions which could artificially lower the estimate of A$_V$ in the regions closer to the source is possible. We also note that from -1.6 to -0.85 arcsec the flux ratios have higher errors than near the jet source. In any case, the A$_V=2^m-6^m$ values measured from the two [\ion{Fe}{II}] ratios are similar to those estimated from H$_2$ (A$_V=1^m-6^m$).

Although the [\ion{Fe}{II}] 1.644/1.257 ratio is mostly lower than the intrinsic ratio, we note that the behaviour of the two ratios correlates throughout the jet. For example, both ratios show local maxima at 1\arcsec, which roughly coincides with the position of the knots in each lobe. It is especially prominent in the [\ion{Fe}{II}] ratios between -1.6 and -0.85 arcsec. The lack of reliable values for A$_V$ in the blue lobe probably means a lower extinction here compared to the red lobe. We also compared the results with the models of \citet{Har04}, which predict the flux ratios of some bright [\ion{Fe}{II}] lines. For example, the models for [\ion{Fe}{II}] 1.644/1.257 and 1.644/1.321 give the same ratio values of 0.96 and 3.70 for all models considered in that work. However, our derived [\ion{Fe}{II}] 1.644/1.257 and 1.644/1.321 ratios are also mostly lower than those predicted by Hartigan models.

\subsection{Emission photocentre shifts}
\label{xy_shift}
To study the gas extension and shape of the emitting region, we analysed the position of the emission photocentre with respect to the stellar continuum. First, Fig.~\ref{lines9} (and Fig.~\ref{Feii_14}) reveals a clear displacement of the maximum of [\ion{Fe}{II}] emission, relative to the stellar position, directly from the observed maps. 

\begin{figure*}
        \centering
        \resizebox{\hsize}{!}{\includegraphics{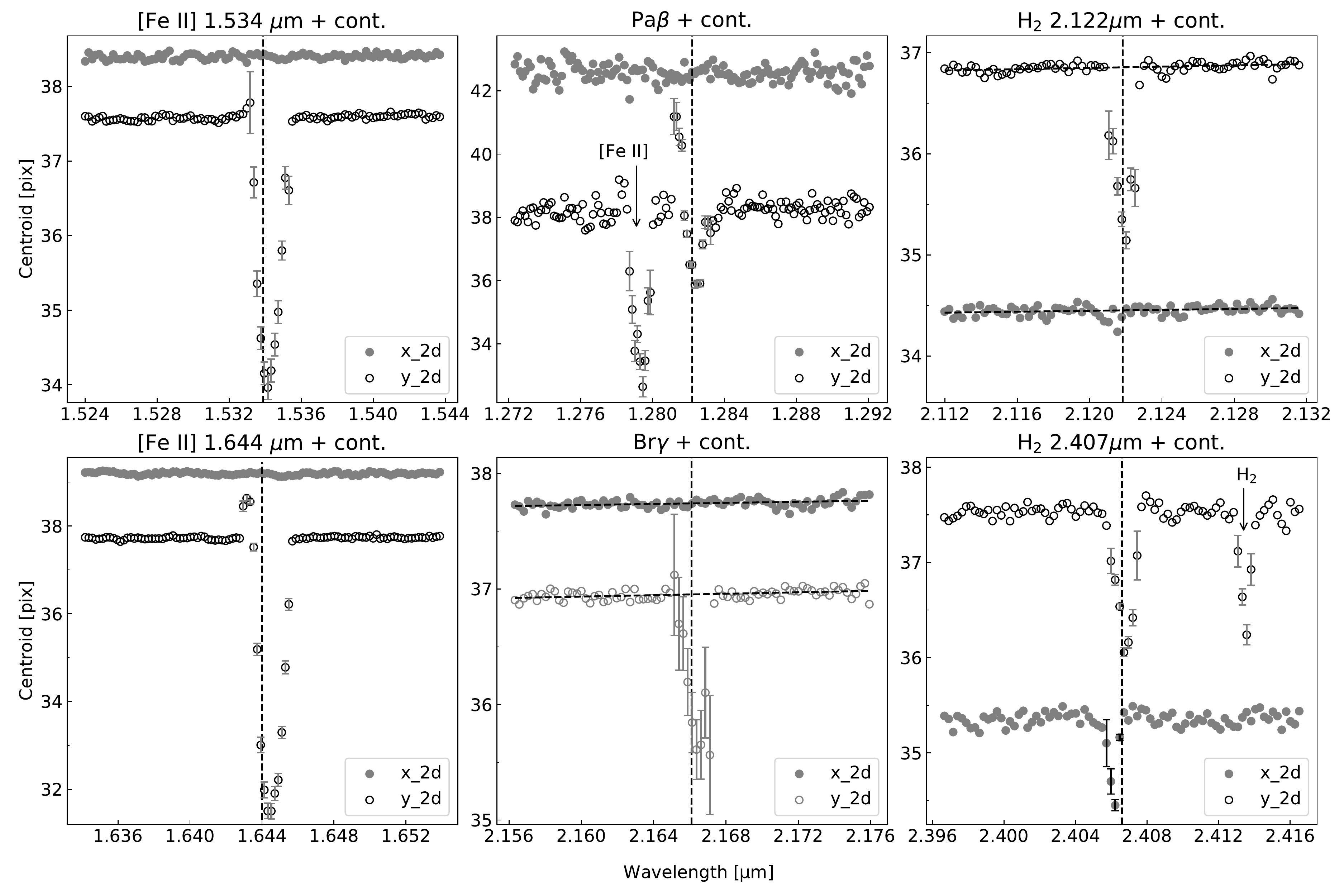}}
        \caption{Photocentre shift for [\ion{Fe}{II}] 1.534, 1.644 $\mu$m, Pa$\beta$, Br$\gamma$, and H$_2$ 2.122, 2.406 $\mu$m along the \textit{XY}-directions. The \textit{Y}-direction is aligned with the jet axis, whereas the \textit{X}-direction is orthogonal. The wavelength of spectral lines in the vacuum is indicated by vertical lines. On some plots, the values of centre coordinates are shifted as a whole in order to place both distributions on the same plot. The instrumental bias around Br$\gamma$, calculated from the continuum, was fit with a linear function (shown as dashed lines) and removed from further analysis. The shifts in the line cores were corrected for the line-to-continuum ratio, and the errors derived for the line centre displacement are shown. All the emission lines display a significant photocentre shift along the \textit{Y}-direction (the direction of the jet), whereas the H$_2$ lines also show a small shift in the orthogonal direction (\textit{X}).}
        \label{xy_pos}
\end{figure*}

To extract the exact $XY$ positions, namely orthogonal and along the jet axis, of the line emission maxima and those of the continuum centre, we used a 2D Moffat fitting, which shows smaller residuals than the 2D Gaussian fitting. To calculate the relative position of the photocentre and avoid an incorrect displacement of gaseous emission with respect to stellar position because of problems with continuum subtraction, we fitted the gaseous emission plus the continuum for these lines. We corrected the derived displacement for the line-to-continuum ratio. Since the signal from the line approaches zero in the wings, the noise (and hence a measured error of the signal) increases exponentially. The effects of background noise can also be considerable; to account for this, we applied a correction to those positions in the line cores where the detection of the line signal was reliable  (S/N$>$0.1; see Fig.~\ref{rv_h2}) and the background noise was low. The derived errors are shown only for these positions. Figure~\ref{xy_pos} shows how the position of the photocenter changes for several bright lines that trace different gaseous structures: [\ion{Fe}{II}] 1.534, 1.644 $\mu$m, Pa$\beta$, Br$\gamma$, and H$_2$ 2.122, 2.406 $\mu$m. To verify these results, we repeated the Gaussian fitting on images with the pure emission after the stellar continuum subtraction. The second method showed that all of the emission lines exhibit a photocentre shift with amplitudes similar to those found in the case of the line plus continuum images.

Our calculated $XY$ distribution of the maximum [\ion{Fe}{II}] emission as a function of wavelength (Fig.~\ref{xy_pos}) reveals that the shift of the forbidden emission with respect to the continuum position follows the jet beams, which the emission traces (i.e. for the blueshifted lobe) and the photocentre shifts to the same direction, while for the redshifted lobe the photocentre shifts along the red lobe direction. Furthermore, the photocentre shift along $Y$ in the redshifted lobe is larger than in the blueshifted direction. This behaviour correlates with the brightness asymmetry of the jet lobes. Our measurements show that the [\ion{Fe}{II}] line at 1.644 $\mu$m, formed with an upper level ($\mathrm{a}^4\mathrm{D}_{7/2}$), and [\ion{Fe}{II}] 1.279~$\mu$m line, formed with $\mathrm{a}^4\mathrm{D}_{3/2}$, reveal similar photocentre shifts in the red lobe (0\farcs31 and 0\farcs28, respectively). At the same time, the [\ion{Fe}{II}] 1.534~$\mu$m lines formed with $\mathrm{a}^4\mathrm{D}_{5/2}$ are displaced by a smaller distance (0\farcs18). At the adopted distance for Th~28 of 185~pc, this means that the [\ion{Fe}{II}] emission with different excitation reaches its maximum at different distances: $\sim$55 and $\sim$33 au from the central star, respectively. At the same time, the strong [\ion{Fe}{II}] 1.534 and 1.644~$\mu$m lines exhibit a photocentre shift of 0\farcs01--0\farcs05 for the blueshifted lobe. The strongest line, [\ion{Fe}{II}] 1.257~$\mu$m, has the strongest shift in the $Y$-direction (0\farcs14).

Unlike the [\ion{Fe}{II}] lines, the emission regions of Pa$\beta$ and Br$\gamma$, as well as the H$_2$ lines, do not show any prominent photocentre shift in the emission maps (Fig.~\ref{lines9}). However, photocentre shifts are clearly detected in the fits (Fig.~\ref{xy_pos}).  For Pa$\beta$ and Br$\gamma$, our calculation shows that this emission displacement along the $X$-direction is not detected, while it is large in the $Y$-direction (along the jet axis), the same for the [\ion{Fe}{II}] lines. The position of the continuum photocentre is quite flat for the Pa$\beta$ and the H$_2$ line at 2.406 $\mu$m, but for Br$\gamma$ and H$_2$ at 2.122 $\mu$m there was a noticeable slope, which was removed by linear fitting during the calculation. In general, the photocentre on the Br$\gamma$ line demonstrates the same behaviour as the [\ion{Fe}{II}] lines: the photocentre has a small shift ($\sim0\farcs02$) for the blueshifted lobe (upper jet lobe in Fig.~\ref{lines9}), but a much larger shift in the longer red jet lobe. Our fitting shows a displacement of 0\farcs07 ($\sim$13 au) for the Br$\gamma$ line in the red lobe. At the same time, after the line-to-continuum correction, the Pa$\beta$ line shows almost the same shift in both jet directions (0\farcs15 for the blueshifted lobe and 0\farcs13 for the redshifted lobe). These results indicate that the gaseous structures traced by the atomic lines are probably affected in the regions where the jet forms and that the atomic gas is entrained by the inner jet.

We also fit the circular cores of the spatial profiles for the bright H$_2$ lines at 2.122 and 2.406~$\mu$m centred on the star and calculated the photocentre displacement with respect to the continuum, also shown in Fig.~\ref{xy_pos}. Both lines exhibit a slight shift along the jet axis ($Y$-direction). Their calculated centroid shows a similar shift (1.5 and 1.7 pix), which corresponds to an average angular shift of $\sim$0\farcs08 ($\sim$15 au). At the same time, both lines also reveal a small displacement in the disc plane ($X$-direction). We took into account the line-to-continuum ratio only for the $X$-shifts of the H$_2$ 2.406~$\mu$m line, because the $X$-shift of H$_2$ 2.122~$\mu$m line is too small for this correction. Both lines show a slight redshifted $Y$-displacement, whereas their $X$-offset is blueshifted and weaker than in the $Y$-direction. The width of the H$_2$ displacement is smaller than that of the [\ion{Fe}{II}] lines and the atomic hydrogen lines. After line-to-continuum correction, the computed photocentre shift in the $Y$-direction of the faint H$_2$ 2.413~$\mu$m line (Fig.~\ref{xy_pos}) has an amplitude similar to the brighter H$_2$ lines ($\sim$0\farcs06 or $\sim$11 au). However, unlike the brighter H$_2$ lines, the H$_2$ 2.413~$\mu$m line does not exhibit any displacement in the $X$-direction.

To summarise, all bright emission lines reveal that the maximum gaseous emission is shifted along the redshifted jet, with the largest shift seen in the [\ion{Fe}{II}] lines, tracing the jet beams. Both the atomic and molecular hydrogen lines also have detectable shifts in the same direction as the red lobe of the jet. The photocentre displacements of the Pa$\beta$ and Br$\gamma$ emission regions, which are usually small and unresolved, imply that the HI emission traces the material entrained by the jet or is due to atomic gas from the jet themselves. These results also reveal a prominent asymmetric spectroastrometric signal in the redshifted jet, compared to the (slower) blueshifted jet.

\subsection{Morphology of the H$_2$ emission}   
\label{h2_region}
Unlike the [\ion{Fe}{II}] emission, which traces the collimated jet lobes, the spatially resolved emission in the H$_2$ lines shows a flattened arc-like structure, orthogonal to the direction of the jet axis. In the brightest lines this H$_2$ emission covers the entire FoV in the north--south direction (see Fig.~\ref{lines9} and Fig.~\ref{lines3}). However, this region may be even larger because the H$_2$ emission at 2.12~$\mu$m, for instance, is still bright (Fig.~\ref{lines9}) at the edges of the SINFONI FoV ($3\arcsec\times3\arcsec$, $\sim540$~au $\times$ 540~au at the adopted distance of Th~28). Therefore, the H$_2$ emission has a size of at least 540~au. If we exclude the region of faint emission at the periphery, the emission around the central star has a bright inner core, with a transverse width of about 1\farcs5 ($\sim$135 au). 
\begin{figure}
        \centering
        \resizebox{\hsize}{!}{\includegraphics{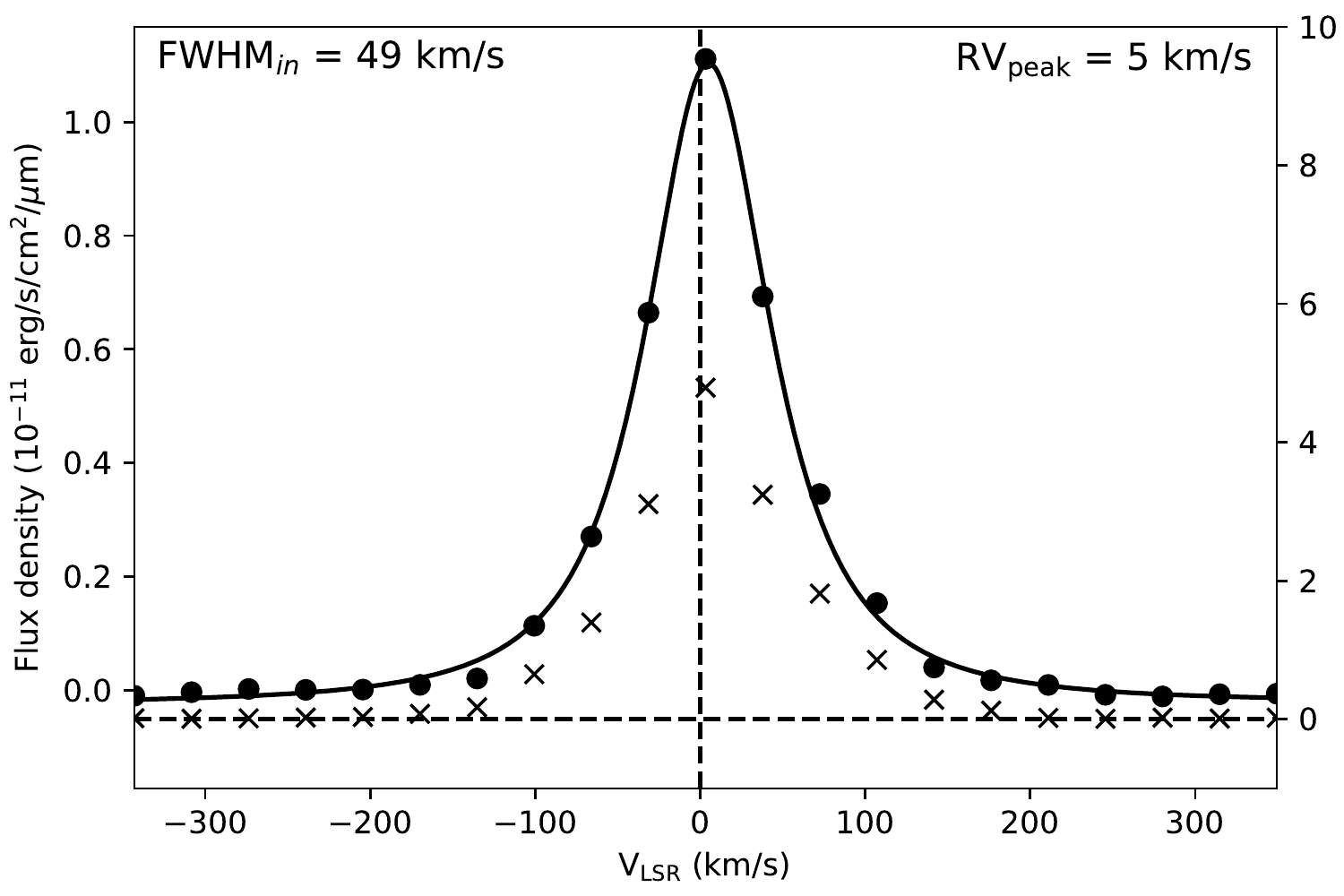}}
        \caption{Fit of the line profile of the H$_2$ 2.122~$\mu$m line (FWHM$_{in}=49\pm 11$ km s$^{-1}$) integrated over the central region using a Moffat function (\textit{solid black line}). The peak of the emission profile is redshifted by $\sim5\pm2.7$ km~s$^{-1}$ with respect to the system velocity of $\sim3$ km~s$^{-1}$ (\textit{dashed vertical line}). The S/N computed for each position is indicated with a \textit{cross}. The S/N scale is shown on the right y-axis, and the \textit{dashed horizontal line} indicates the zero level.}
        \label{rv_h2}
\end{figure}

\begin{figure*}
        \begin{minipage}[c]{4.3cm}
        \resizebox{\hsize}{!}{\includegraphics{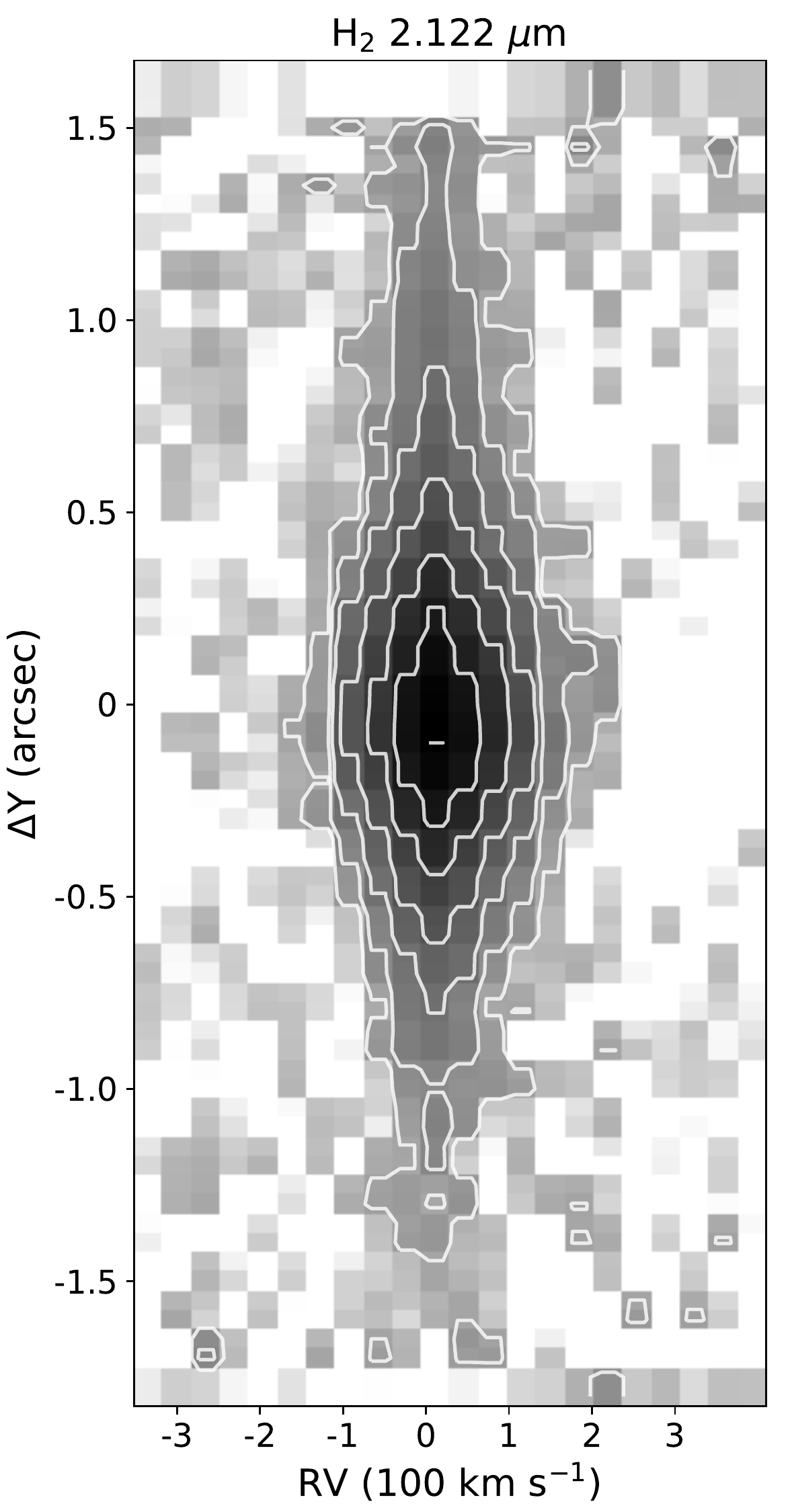}}
\end{minipage}
\begin{minipage}[c]{2.75cm}
        \resizebox{\hsize}{!}{\includegraphics{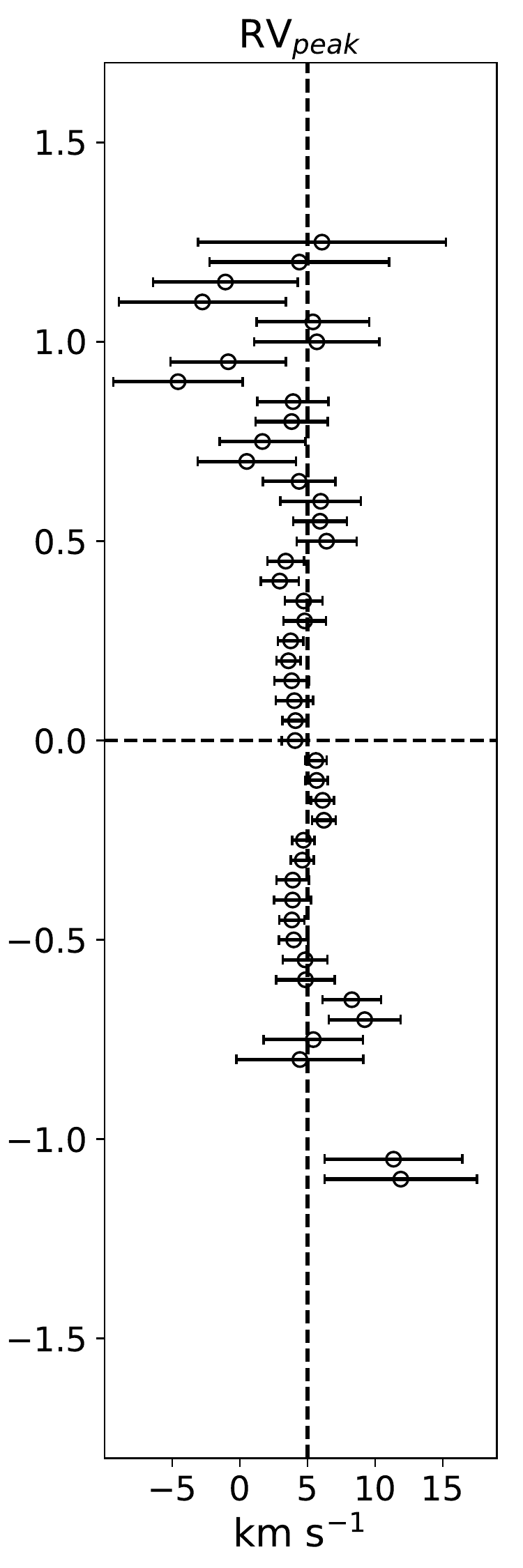}}
\end{minipage}
\begin{minipage}[c]{2.75cm}
        \resizebox{\hsize}{!}{\includegraphics{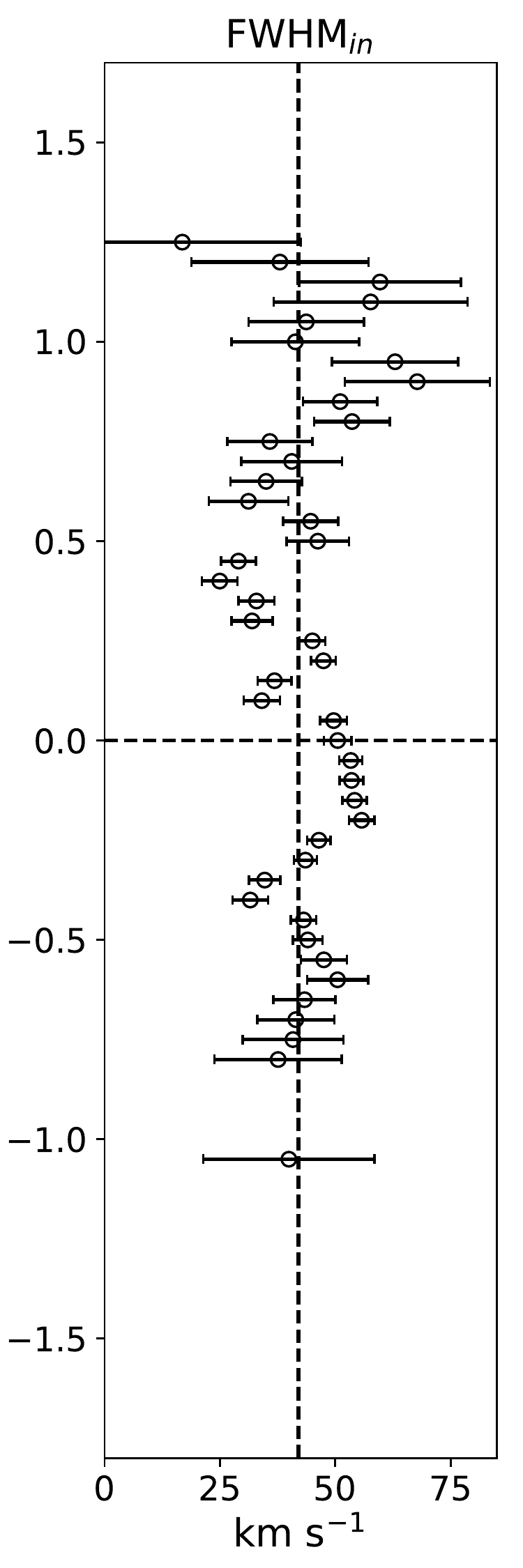}}
\end{minipage}
\begin{minipage}[c]{8.45cm}
        \resizebox{\hsize}{!}{\includegraphics{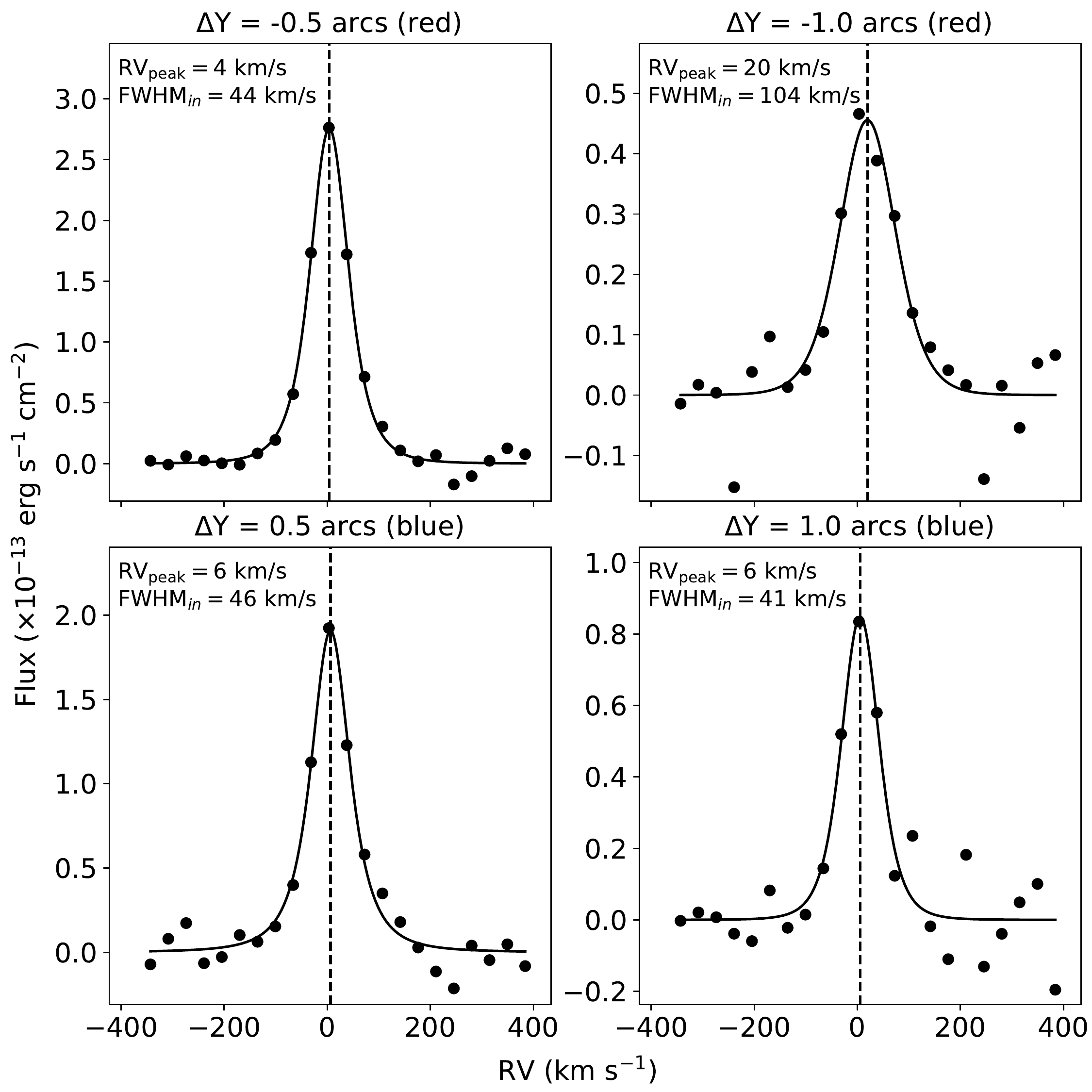}}
\end{minipage}
        \caption{Distribution of the H$_2$ 2.122 $\mu$m emission along the jet. \textit{Left image}: PV diagram of the H$_2$ 2.122 $\mu$m line along the jet axis of Th~28, obtained by integrating the fluxes in the 22 spectral frames over the direction perpendicular to the jet. Logarithmic contours are drawn from $2\times10^{-14}$ erg s$^{-1}$ cm$^{-2}$ arcsec$^{-2}$ \AA$^{-1}$, with an incremental logarithmic step of 0.3 dex. \textit{Two central panels} show the derived RV$_{peak}$ and FWHM$_{in}$ as a function of distance from the source. A vertical line at 5 km s$^{-1}$ in the RV panel marks possible systemic RV shift for the line;  an average value of FWHM$_{in}$ is 43 km s$^{-1}$ (\textit{vertical line}). \textit{Right four panels}: Spectral profiles across the jet at selected positions within an $\pm 1\arcsec$ area in the blue and red lobes (\textit{black points}). The data were fit using a Moffat function (\textit{solid line}); the derived peak RV and intrinsic FWHM are shown at the top of each panel.}
        \label{pv_h2_jet}
\end{figure*}

Our fitting of this core in several bright H$_2$ lines shows that the extension of this core in the $X$-direction (disc plane) is slightly larger than in the $Y$-direction (jet axis) for all the fitted lines. For instance, at 2.122~$\mu$m H$_2$ has FWHM$_X$=0\farcs383, but FWHM$_Y$ is 0\farcs372. The difference between $X$ and $Y$ is small ($\sim$3\%), but visible in almost all H$_2$ lines;  even after excluding the faint emission periphery, the central H$_2$ core has a flattened shape aligned with the plane of faint peripheral emission and therefore with the circumstellar disc. The FWHM of the continuum around the H$_2$ 2.122~$\mu$m line is the same in both directions (i.e. FWHM$_X$=FWHM$_Y$=0\farcs325), and it is smaller than that of the line centre. We calculated the total flux against the wavelength for three bright molecular lines in the $K$ band, H$_2$ 1.958, 2.122, and 2.413~$\mu$m, and fit the 1D emission line profiles to calculate their FWHMs. The fitted FWHM$_{in}$ of H$_2$ 2.122 is $\sim$50 km~s$^{-1}$ (Fig.~\ref{rv_h2}), which is narrower than Br$\gamma$ and Pa$\beta$ (internal FWHM$_{RV}$(H$_2$ 1.958)=89 km~s$^{-1}$ and FWHM$_{RV}$(H$_2$ 2.407)=43 km~s$^{-1}$). The morphology of the  H$_2$ emission (Fig.~\ref{lines9}) differs from that of [\ion{Fe}{II}], which clearly traces the outflow structure. The fitted peak RV for H$_2$ 2.122~$\mu$m is centred at 5 km~s$^{-1}$, which differs from the result of \citet{Cof10} who found a blueshifted velocity of about $-10$ km s$^{-1}$ for the H$_2$ emission in the direction of both the receding and approaching jet lobes. We try to explain the H$_2$ RV shift analysing the H$_2$ PV diagram along the jet axis (see Fig.~\ref{pv_h2_jet}). We also note that \citet{Cof10} did not resolve the H$_2$ emitting region from the measurement along the jet, while for the slit position perpendicular to the jet, the H$_2$ emission is clearly resolved (see Figs.~3 and 7 in their article). This agrees well with our results from the SINFONI data.

\begin{figure}
        \centering
        \begin{minipage}[c]{9cm}
                \resizebox{\hsize}{!}{\includegraphics{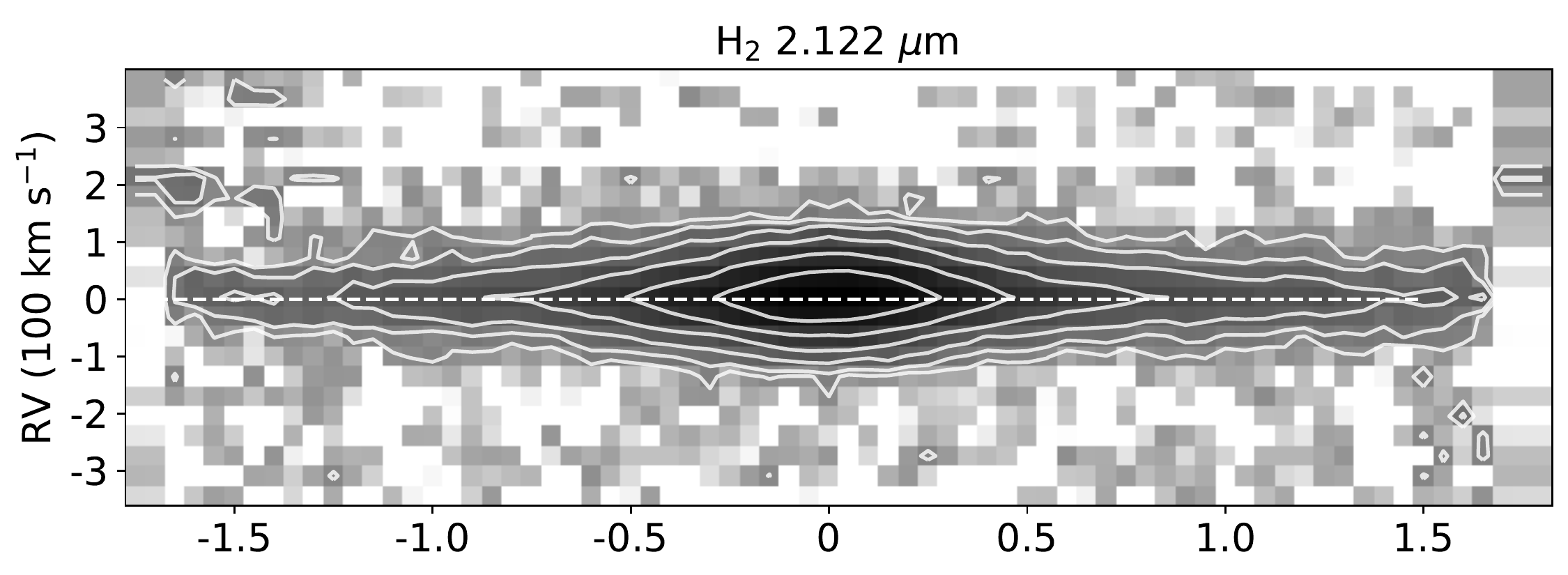}}
        \end{minipage}
        
        \vspace{-3pt}
        \begin{minipage}[c]{9cm}
                \resizebox{\hsize}{!}{\includegraphics{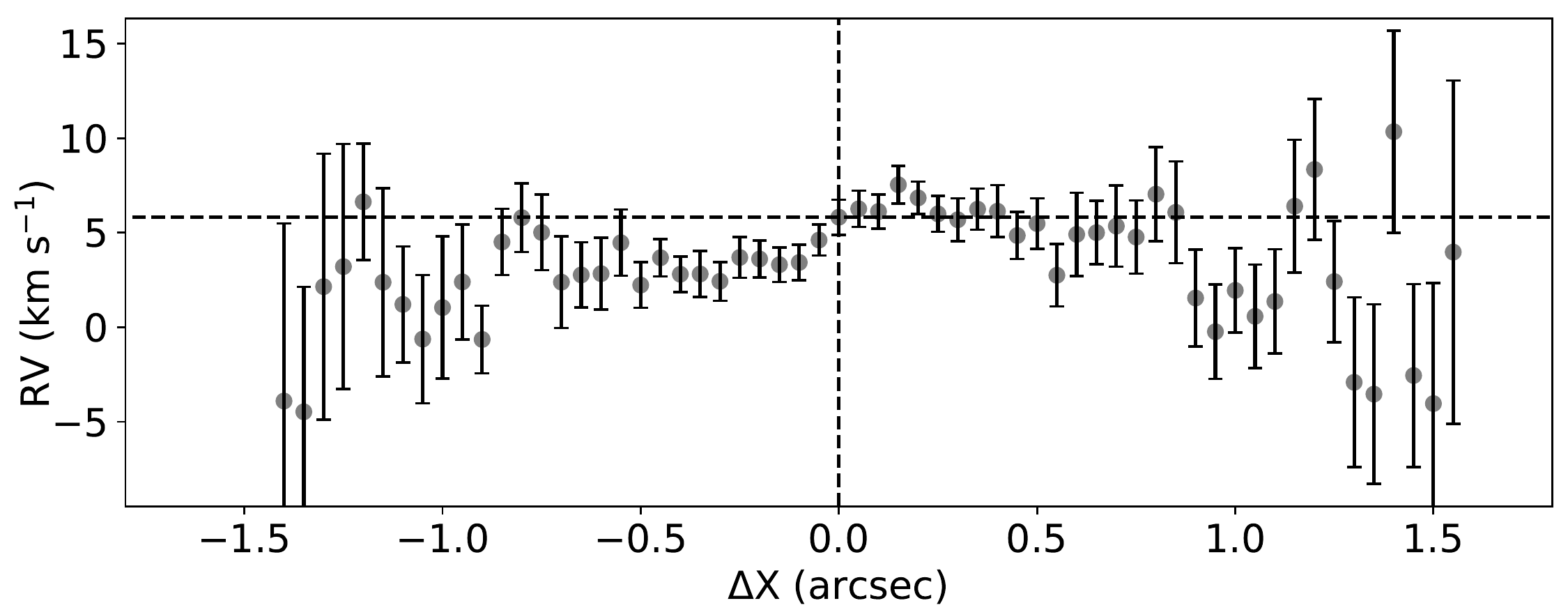}}
        \end{minipage}
        \caption{Distribution of the H$_2$ 2.122 $\mu$m emission in the disc plane. \textit{Upper panel}: PV diagram of the H$_2$ 2.122 $\mu$m line perpendicular to the direction of the jet axis and in the disc plane. \textit{Lower panel}: Peak radial velocities with error bars  computed from fitting across the PV diagram  at the every position with 0\farcs05 steps. The region from $-0\farcs75$ to $0\farcs75$ around position at $0\arcsec$ corresponds to the bright core of H$_2$ emission around the jet source.}
        \label{pv_h2_disc}
\end{figure}

Using our SINFONI observations, we calculated the PV maps of the H$_2$ 2.122~$\mu$m line both along (Fig.~\ref{pv_h2_jet}) and perpendicular to (Fig.~\ref{pv_h2_disc}) the direction of the jet to explore the kinematics of molecular hydrogen emission. Although our 2D images show that H$_2$ emission is concentrated mainly in the plane orthogonal to the jet axis, the PV map along the jet axis shows hydrogen emission up to $\pm1\farcs5$ away from the jet source. This scale is about twice the radius of the bright compact core. The H$_2$ 2.122~$\mu$m emission distribution along the jet axis is, in general, similar to that seen on the more extended PV map \citep[Fig.~2 from][]{Cof10}, which traces H$_2$ emission up to $\sim2\arcsec$ in both jet lobes. The H$_2$ emission does not show a brightness asymmetry between the blue and red lobes, as seen in the forbidden lines, which also agrees well with the result of \citet{Cof10}. The distribution of RV$_{peak}$ and FWHM$_{in}$ along the jet derived from fitting the PV diagram are shown as a function of distance from the source. Fitting the H$_2$ emission profiles has smaller errors for the distance from the jet source $<0\farcs7$, but the fitting errors increase at larger distance from the source due to the poorer S/N of the line (Fig.~\ref{pv_h2_jet}). In general, the PV diagram does not show a significant difference between the H$_2$ gas velocity along the jet at distances $>0\farcs5$ compared to the internal regions $<0\farcs5$. The average values of the parameters computed at all positions with reliable fitting are RV$_{peak}=4.6\pm 1.9$ km s$^{-1}$ and FWHM$_{in}=42.4\pm 4.7$ km s$^{-1}$ and both values are marked in Fig.~\ref{pv_h2_jet} by vertical lines. The distribution of RV$_{peak}$ values along the jet is very close to the mean value, and there is no significant difference between the red and blue lobes. Therefore, the RV$_{peak} \sim 5$ km s$^{-1}$ visible in other RV diagrams (Fig. \ref{rv_h2} and \ref{pv_h2_disc}) probably represents a systemic shift resulting from errors in the wavelength calibration. However, we can draw some conclusions. In particular, the rest RV of the H$_2$ emission in the jet lobes is very small, and therefore the total H$_2$ velocity in the jet should also be small. There are a few points at distant positions where the RV sign agrees with the sense of the inclination of the jet lobes (i.e. H$_2$ 2.122~$\mu$m has the negative RV for the approaching lobe and the positive values for the receding one). However, the measurement errors for these positions are also high.

The PV diagram of the H$_2$ line along the direction perpendicular to the jet axis is also similar to that of \citet{Cof10} (see their Fig.~7). The horizontal dashed line shows the RV of the H$_2$ gas derived at the stellar position (5.4 km s$^{-1}$). The size of the region from $-0\farcs75$ to $0\farcs75$ corresponds to the bright core of H$_2$ emission around the jet source. The RV uncertainties are quite small for this central region and then increase towards the periphery of the H$_2$ emission region. If we consider the bright emission region, we note that the left part (from $-0\farcs75$ to $0\arcsec$) shows a systematically lower RV value than the opposite one (from $0\arcsec$ to $0\farcs75$) which can be an indication of gas rotation in a circumstellar disc. Both the direction and the magnitude of the velocities are similar to that derived from CO emissions that was interpreted as the rotation of the circumstellar disc \citep{Lou16}. However, the current spectroscopic resolution does not allow us to draw any conclusions about the Keplerian rotation in the disc. Taking the error bars into account, the PV map does not show a large departure of RVs along the disc plane direction from the velocity at $0\arcsec$. The average FWHM$_{in}=47.3\pm 4.0$ km s$^{-1}$ within $-0\farcs75$ to $0\farcs75$ along the disc plane is slightly higher than the value measured along the jet. The RVs observed in the faint peripheral H$_2$ emission (at radial distances $>0\farcs75$) seem to deviate from that of the central region and are consistent with zero, but the measurement errors are higher for those positions as well.

Analysis of the morphology of H$_2$ emission, as well as gas parameters, shows that many YSO outflows and jets are associated with molecular hydrogen emission generated by gas shock waves \citep[][and references therein]{Beck08}. However, some YSOs also show the presence of quiescent molecular hydrogen emission, probably of fluorescent origin \citep{Beck08}. Therefore, the origin of the H$_2$ emission requires special consideration, which we discuss in Sect.~\ref{discuss}. 

\subsection{Physical parameters of the H$_2$ emission region}
\label{h2_param}

The H$_2$ transitions that arise from the same upper state of the H$_2$ molecule can be used for the determination of $A_V$. For example, bright H$_2$ lines such as the $\nu=1-0$~S(1) at 2.122~$\mu$m and $\nu=1-0$~Q(3) at 2.424~$\mu$m are both detected in the spectrum of Th~28. The intrinsic Q(3)/S(1) line ratio of 0.7 is not affected by the physical conditions of the gas-emitting environment. At the same time, $A_V$ calculations based on the H$_2$ line ratios have a very strong dependence on the values of the ratio \citep[e.g. ][]{Dav11}. Taking the integral fluxes of these lines within a central emission core of $1\arcsec\times1\arcsec$ in size and the relation $A_V = -114 \times \log(0.704 \times [I_\mathrm{S1}/I_\mathrm{Q3}])$ from \citet{Dav11}, we find a mean extinction of $A_V=6\pm0.3$~mag for this region. This result differs from the value of $A_V=1.1$~mag derived from the pre-main-sequence stellar evolutionary tracks of \citet{Fran12}. The last result is based on photometry, probably obtained with a much larger aperture and therefore less affected by absorption in the circumstellar disc.

\begin{figure}
        \centering
        \resizebox{\hsize}{!}{\includegraphics{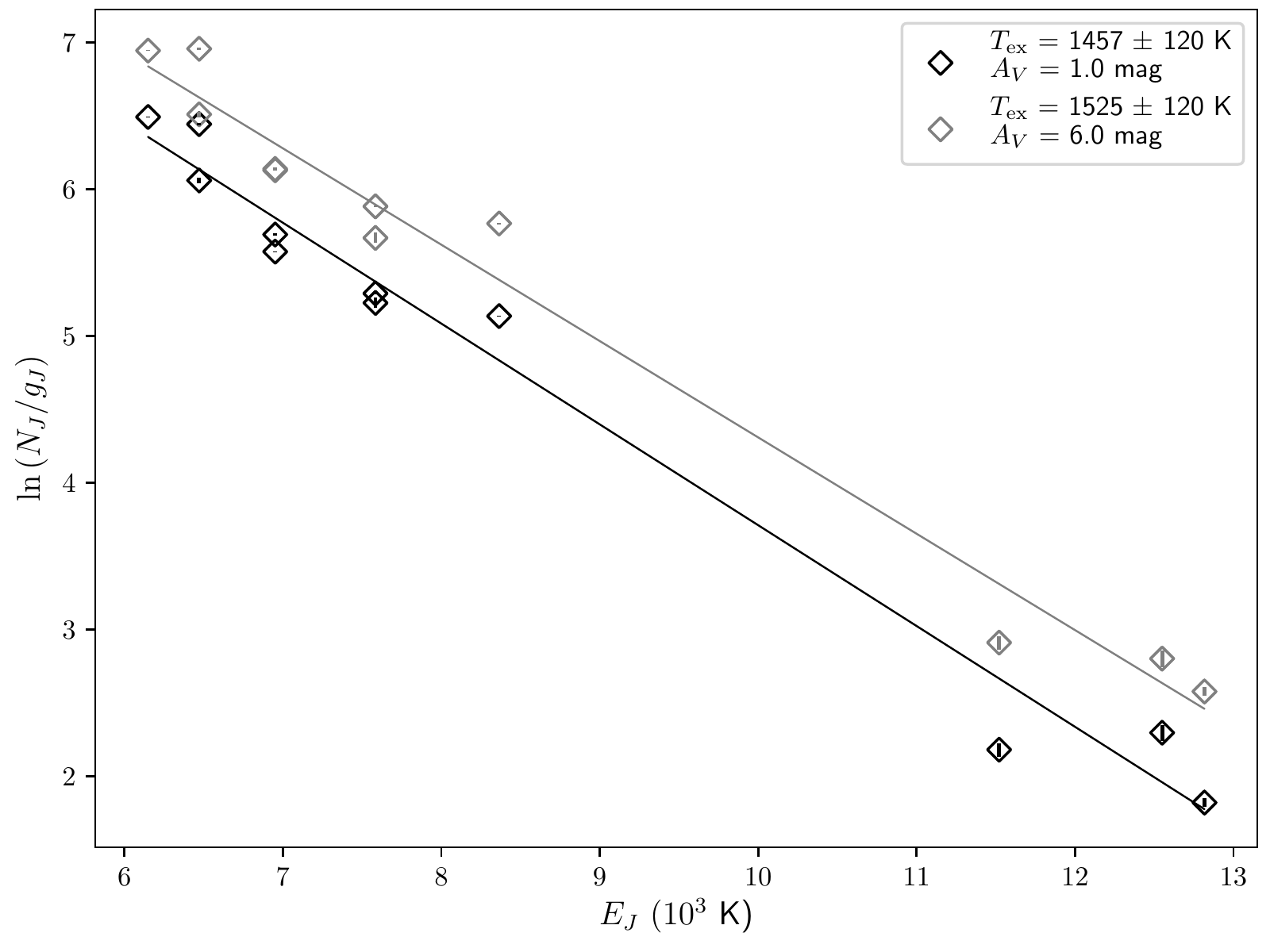}}
        \caption{Rotation diagram for H$_2$ lines in the central ($0\farcs25\times 0\farcs25$) region of Th~28. The \textit{solid lines} show the best fit to the logarithm of column density vs upper energy level. Two colours represent the impact of different values of $A_V$ on the excitation temperature. $A_V$ = 1.0 mag is taken from \citet{Fran12}; $A_V$ = 6.0 mag is derived from the $I_\mathrm{S1}/I_\mathrm{Q3}$ relation.}
        \label{AV}
\end{figure}

\begin{figure}
        \centering
        \resizebox{\hsize}{!}{\includegraphics{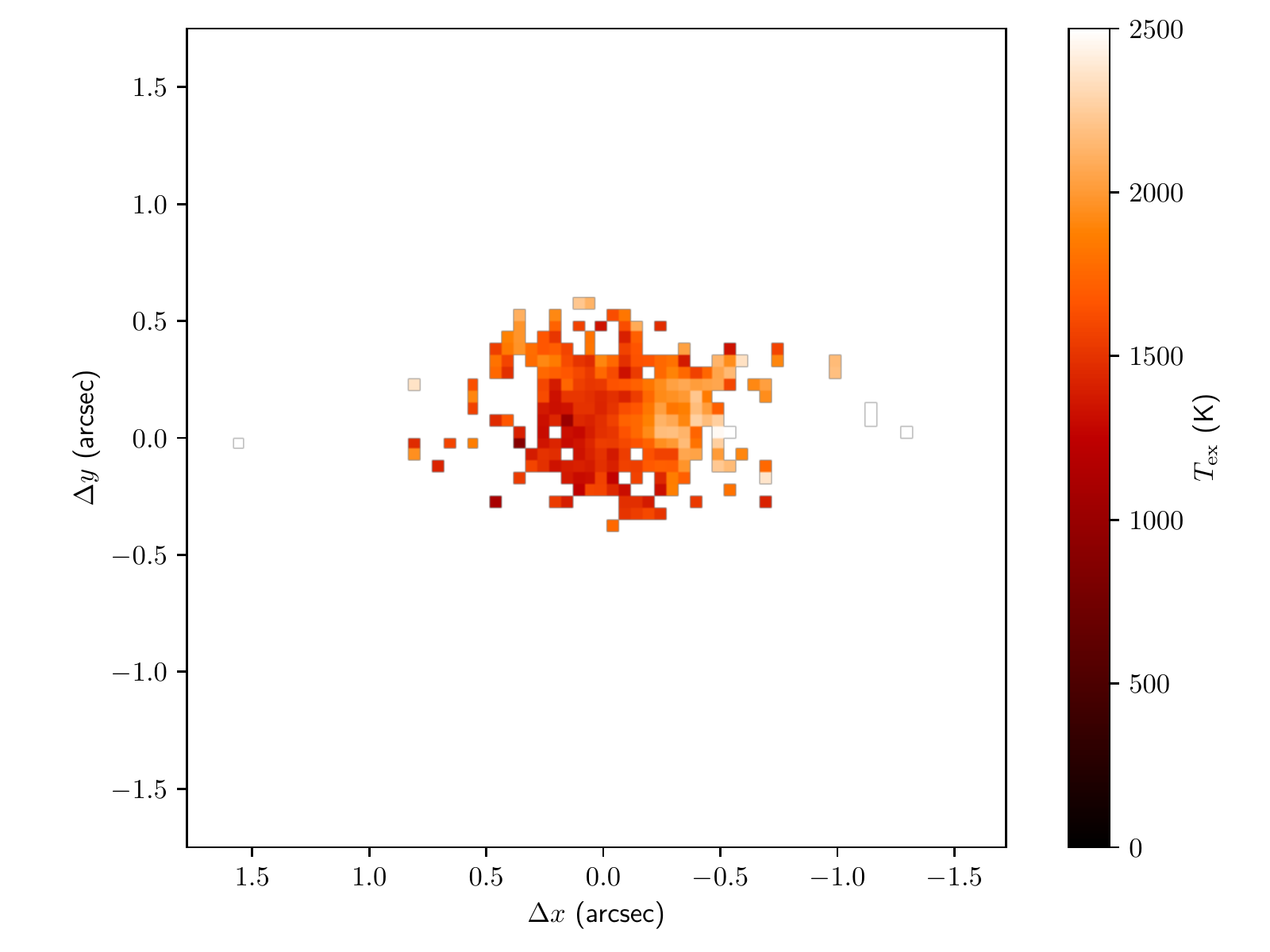}}
        \caption{H$_2$ excitation temperature map. The values from 0 to 2500 K in linear scale are colour-coded from dark to light (see colour scale at right). All data points selected have at least nine molecular hydrogen lines with  flux $>$ 2$\sigma$.}
        \label{temp_h2}
\end{figure}

The excitation temperature of H$_2$ can be estimated using the rotational diagram method. For a Boltzmann distribution between energy levels $N_{J} / g_{J} \sim e^{E_{J}/kT}$, where $N_{J}, g_{J}, E_{J}$, and $T$ are column density, statistical weight of the transition, upper-level energy, and excitation temperature, respectively. This implies that a local thermodynamic equilibrium (LTE) has been established in our system. The column density may also be expressed through flux as $N_J = 2 F_J \lambda_J / A_J \hslash c$, where $F_J$, $\lambda_J$, and $A_J$ are the dereddened flux, line wavelength, and the Einstein coefficient of each transition, respectively. For dereddening, we used the extinction law of \citet{ccm89} with $R_V=3.1$ (see below for an additional discussion on the values of $A_V$). In Fig.~\ref{AV}, we show the dependence of the natural logarithm of the H$_2$ column density of each level on the energy of the upper level for the central region, integrated in $0\farcs25\times 0\farcs25$.  The uncertainties were calculated using the bootstrap method by randomly varying the flux values with a normal distribution over their measured uncertainties and finding the dispersion of the resulting temperatures after optimisation. The temperature map obtained is shown in Fig.~\ref{temp_h2}. Values range approximately from 1\,000~K to 2\,500~K.

This Boltzmann diagram method could also be used as another way to determine interstellar extinction, simultaneously with the evaluation of excitation temperature. In our case, the extinction values obtained were sporadic from pixel to pixel; therefore, we used this method only to determine the excitation temperatures. As seen in Fig.~\ref{AV}, extinction values affect the temperature only within the error margin, which is why the temperature can be determined under the assumption that the excitation is constant.

\section{Discussion}
\label{discuss}

The high-resolution 2D SINFONI spectra allowed us to study the morphology of gas emission formed in different gas structures around Th~28. Three groups of emission lines were detected in NIR spectroscopy: forbidden iron lines, atomic hydrogen lines, and molecular hydrogen lines. Some of our results are similar to those of the detailed study done by \citet{Cof10} based on the analysis of a similar set of emission lines and 1D slit spectra. In that study line fluxes and radial velocities are probably more sensitive because of the long-slit observations, which have higher spectral resolution. However, the 2D SINFONI maps give us a closer look at the spatial distribution of the emission at high angular resolution.

\subsection{Origin of [\ion{Fe}{II}] emission}
\label{orig_fe}

The [\ion{Fe}{II}] lines are emitted by a well-defined outflow, which is extended and coincident with the jet, as traced by [\ion{S}{II}] \citep{Wang09}. The brightness distribution of [\ion{Fe}{II}] along the bipolar jet obtained from the 2D maps confirms the jet asymmetry visible in the [\ion{S}{II}] map, and coincides in general with the results of \citet{Cof10}. In the vicinity of the central source, the [\ion{Fe}{II}] emission is distributed approximately uniformly in the jet column, but forms emission knots on both sides of the source at a distance of $\sim1\arcsec$, which were not detected by \citet{Cof10}. More distant emission knots were also found by \citet{Wang09}, but are outside our observed FoV.

\begin{figure}
        \centering
        \resizebox{\hsize}{!}{\includegraphics{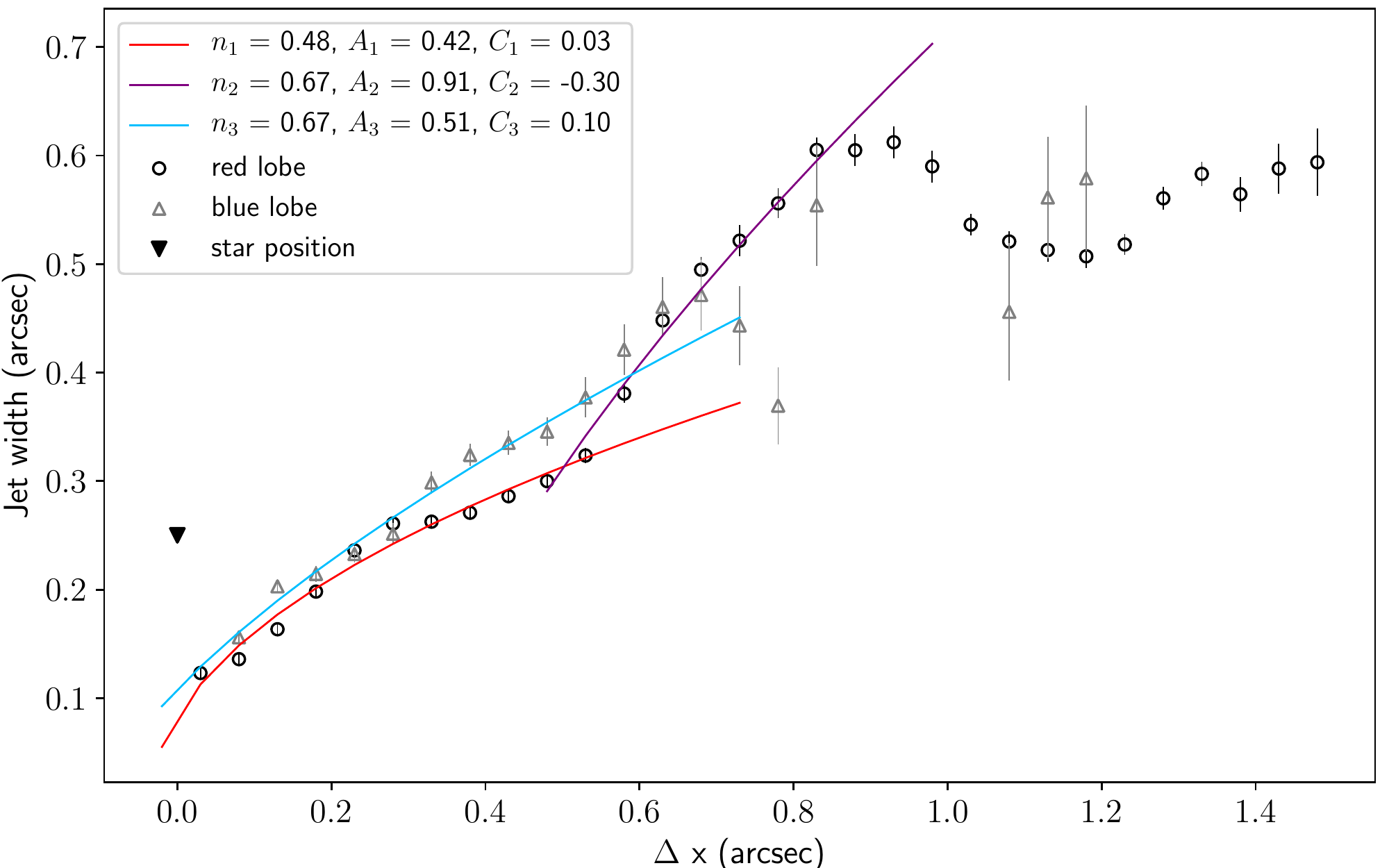}}
        \caption{Jet width vs distance from the central source at 1.644~$\mu$m. The intrinsic jet width is computed with FWHM$_{PSF}=0\farcs23$ derived from the continuum images around the line. Data from the red and blue lobe of the jet are indicated by \textit{grey triangles} and \textit{black circles}, respectively. The \textit{solid lines} are approximation functions $f(x)\propto(x-C)^n$, where C is a free parameter and n is a degree, which demonstrates the change in geometry of the jet for the red lobe. }
        \label{js}
\end{figure}

The emission in the iron lines allows us to trace the change in the geometry of the jet. We integrated the flux of the bright [\ion{Fe}{II}] 1.644 $\mu$m line over all frames from $-220$ km~s$^{-1}$ to $+240$ km~s$^{-1}$. Figure~\ref{js} shows the width of the jet measured using the [\ion{Fe}{II}] 1.644 $\mu$m line as a function of the distance from the central source. The observed FWHM is measured with a Gaussian fit perpendicular to the jet axis. The intrinsic jet width is calculated as the observed FWHM corrected for the FWHM$_{PSF}= 0\farcs23$ derived around the [\ion{Fe}{II}] line (i.e. FWHM$^2_{jet}$=FWHM$^2_{[\ion{Fe}{II}]}$-FWHM$^2_{PSF}$). Data were masked at a level of $3\sigma$. To estimate the size of the jet cone at its base, we fit each of the lobes with a power function $f(x)\propto(x-C)^n$, which are shown in the plot. These approximations reveal a change in the geometry of the red jet lobe at about 0\farcs5 from the central source. A similar change is seen in the blue lobe of the jet, but there is already a significant deformation of the jet shape at 0\farcs6, probably due to the disc--jet interaction, and therefore the fit is not applicable. At around 0\farcs9, however, the collimation of the red jet lobe seems to change again and the diameter of the jet stops growing and becomes more stable. A similar behaviour is detected in some other YSO jets; the width of the jets first increases with increasing distance from the jet source, but at certain distances it stops growing and its collimation stabilises \citep[e.g. the width of the RW Aur jet stabilises at $\sim$1\farcs5 from its source;][]{Woi02}. Therefore, at larger distances the Th~28 jet is found to be more collimated, as measured by \citet{Mur21}.

Using the approximation function for the jet width, an upper limit to the maximum possible size for the jet formation region is now available by calculating the diameter of this function at the level of the accretion disc. The maximum radius of the jet at the disc position is estimated to be $\sim$0\farcs015, corresponding to a jet launching radius of $\sim$3 au. This estimate represents an upper limit of this value due to the high uncertainty of the extrapolation and is higher than that derived by \citet{Fer06}. At the same time, this is less than the maximum launch radius (r$_{max}\simeq 5$ au) adopted in the magnetohydrodynamic jet models of \citet{Ray07}.

We used a simple linear fit to estimate the full opening angle for both outflow lobes in Th~28. For the full range of iron emission velocities, we found that within the first 0\farcs6 the opening angle is $\sim$28\degr$\pm$2\degr\ for both lobes, while the red lobe within 0\farcs9 shows a mean opening angle value of $\sim$35\degr$\pm$3\degr. The derived opening angle also depends on the velocity range. For example, \citet{Agra11} found that the medium velocity gas ($<150$ km~s$^{-1}$) in the DG Tau jet is collimated inside a cone with a full opening angle of 28\degr, similar to that of Th~28, whereas the high-velocity gas ($>150$ km~s$^{-1}$) in the jet is more collimated and has a full opening angle of 8\degr. We computed the opening angles in the Th~28 jet for two velocity ranges,  for emission with a velocity $>100$ km~s$^{-1}$ and with a velocity $<100$ km~s$^{-1}$. We found that at high velocities, the gas in the blue lobe appears to be more collimated with a full opening angle of 20\degr$\pm$3\degr, which increases to $\sim$30\degr$\pm$4\degr\ at low velocities. At the same time, in the red lobe the gas emission has a similar collimation ($\sim27$\degr$\pm$4\degr) for both velocity ranges. Therefore, the Th~28 jet appears to be less collimated than some other YSO jets, with an opening angle of 5--10\degr\ near the jet source \citep{Bac99b,Woi02,Dou04}. At the same time, the DG Tau jet \citep{Woi02} also shows a steep increase in jet width, which is similar to the Th~28 red lobe case.

An interesting feature of the [\ion{Fe}{II}] RV distribution is the presence of opposite motion for both jet lobes: channel maps exhibit a blueshifted emission gas for the redshifted lobe, and vice versa. Taking into account the estimated full opening angle (see the discussion of Fig.~\ref{js}), a natural explanation for this seemingly contradictory result is that the jet cone has a position on its axis close to the sky plane. In this case, we see various velocity components of diverging streams and/or transverse motion resulting from rotation. 
With the adopted inclination of the jet axis and a half-opening angle of 14\degr\ obtained within the first 0\farcs6, the outflow cone would remain completely behind the plane of the sky. Therefore, this opening angle would not be large enough to explain the blueshifted emission in the redshifted flow (Sect.~\ref{jet_kin}). However, with a half-opening angle of $\sim$18\degr\  the outflow cone would reach the front side of the plane of the sky and create a quite prominent negative velocity flow component visible between 0\farcs5 and 1\farcs5 in the red jet lobe.

Various magnetohydrodynamic jet-launching models \citep[e.g.][]{Fer97,Fer06,Pud07} predict the presence of gas flows with different velocities close to the source position, including low-velocity uncollimated outer streams. For example, \citet{Gar08} found that the atomic and molecular gas (seen in [\ion{Fe}{II}] and H$_2$ lines) at the base of HH~34 jet show two components at high velocities (HVC) and low velocities (LVC), which trace the jet and disc wind, respectively. Due to the high inclination of the jet in Th~28, the measured RVs of the [\ion{Fe}{II}] jets are also relatively small, and therefore these lobes are only weakly spectrally resolved in the channel maps. We did not find any indication of low or high components in the redshifted jet. \citet{Cof10} also did not report any distinction between the high- and low-velocity components. However, the fact that RV increases along the blue lobe may imply that both LVC and HVC could be present at the base of the jet. The emission could be dominated by the LVC at the jet base and the HVC further away. This is also supported by the fact that in the blue jet the [\ion{Fe}{II}] line widths are substantially larger at the base of the flow, suggesting the presence of an additional component of the HVC. This is also supported by the presence of [\ion{O}{III}] $\lambda$ 5007 \AA\  in the blue lobe, as reported by \citet{Com10}, who associated this emission with stellar winds.

In general, the [\ion{Fe}{II}] emission at 1.644~$\mu$m covers redshifted velocities up to $\sim$240~km~s$^{-1}$, which is similar to that of the opposite lobe (RV$_{max}\sim$220~km~sec$^{-1}$). From the fit to the line profiles, we adopted RV$_{peak} =41\pm0.8$ km~s$^{-1}$ and $-105\pm7.5$ km~s$^{-1}$ at +1\arcsec\ for the red- and blueshifted lobes, respectively (see Fig~\ref{pv_feii}). This difference is consistent with previous studies, which reported that RVs of HH objects in the blueshifted jet are greater by an average factor of 1.5--2 compared to those in the redshifted jet \citep{Com10,Mur21}. If we adopt the different inclination angles for the opposite jet lobes measured by \citet{Mur21} ($i_{red}=83\degr.5$ and $i_{blue}=77\degr$), the deprojected gas velocity for the emission will be similar for the two beams,   $\sim$360 km~s$^{-1}$ for the redshifted lobe and $\sim$470 km~s$^{-1}$ for the blueshifted lobe. The deprojected velocities calculated from the velocities measured by \citet{Mur21} are $\sim$270 km~s$^{-1}$ for the redshifted lobe and $\sim$360 km~s$^{-1}$ for the blueshifted lobe (i.e. $\sim$100 km~s$^{-1}$ less than our results). For our RVs this difference is $\sim$10 km~s$^{-1}$. We suggest that this RV discrepancy can be partly explained by a systemic shift between these wavelength calibrations. In any case, this deprojected velocity implies that the Th~28 jet is quite fast among the YSO jets because the YSO jets usually show gas velocities between 50 and 400 km~s$^{-1}$ \citep{Fer06}.

One of the interesting features of the data is a prominent brightness asymmetry in the opposite jet lobes, visible in [\ion{Fe}{II}]. One of the possible origins suggested for such an observed jet brightness asymmetry could be an occultation by a circumstellar disc \citep[e.g.][]{Whi14}, which partially blocks the radiation from one of the jet lobes. However, the observed jet structure in Th~28 does not support this scenario. The circumstellar extinction in the direction to Th~28 does not seem to be high. From the analysis of the H$_2$ emission, we estimate the A$_V \approx 6$ mag, while \citet{Fran12} found A$_V = 1.1$ mag based on photometry. This agrees with the fact that we do not see any evidence of a dark lane in Th28, as seen in the DG Tau system \citep[Fig.~1 in][]{Agra14}. Moreover, the redshifted jet is brighter than the blueshifted jet, whereas in the case of disc occultation, we would see the opposite effect. 

Another possibility to explain the observed jet emission structure is the orientation of its natal cloud, where the blueshifted jet lobe penetrates into the dense part of the cloud, while the redshifted jet propagates into a lower-density environment. To verify this, we measured the fluxes for the $\nu=1-0$~Q(3) and $\nu=1-0$~Q(1) H$_2$ lines, and calculated their ratios to the $\nu=1-0$~S(1) line. The first two H$_2$ lines can also be detected along the blue lobes up to 1\farcs5 as H$_2$ 2.122 $\mu$m. At the same time, the bright [\ion{Fe}{II}] 1.644 $\mu$m line in the blue lobe becomes fainter after 0\farcs 5 and disappears completely after 1\arcsec. If the [\ion{Fe}{II}] line weakening is caused by extinction, $I_\mathrm{Q3}/I_\mathrm{S1}$ should also change accordingly. However, the distribution of the H$_2$ ratio is flat and does not show any gradient. The mean value of the H$_2$ ratio for both lobes is $0.79\pm0.21$, which corresponds to $A_V=6$ mag. The $I_\mathrm{Q1}/I_\mathrm{S1}$ distribution along the jet axis is also flat. Therefore, we suggest that the observed brightness asymmetry of the [\ion{Fe}{II}] line has an intrinsic origin, rather than due to changes in extinction along the blue lobe (i.e. the asymmetry is caused by the engine, rather than the medium).

\subsection{Origin of Pa$\beta$ and Br$\gamma$ emission}

The origin of atomic hydrogen emission around the YSOs is still being debated. The theory predicts that the Pa$\beta$ and Br$\gamma$ lines should form close to the star due to accretion. In fact, the luminosity of hydrogen lines has been found to correlate strongly with the accretion luminosity in classical T~Tauri stars \citep{Muz98} and in young brown dwarfs \citep{Nat04}. \citet{Gar08} reported wide Br$\gamma$ emission in the spectrum of HH~34, which, nevertheless, does not show spatial extension along the jet.

However, there is some evidence that this permitted emission can also trace young stellar jet activity \citep{Whe04,Beck10,Car16}. For instance, \citet{Beck10} spatially resolved extended Br$\gamma$ emission in several outflows from young stars, and associated the emission with known HH objects. In the case of the Th~28 jet, \citet{Cof10} concluded from the high-resolution 1D spectroscopy analysis that a component of the Pa$\beta$ and Br$\gamma$ emission comes from the outflow. In the present work, Pa$\beta$ emission is clearly resolved in the redshifted jet lobe. The presence of atomic hydrogen emission in the jet and HH objects is important because these permitted transitions originate in high-density gas regions where forbidden lines are quenched \citep{Cof10}.

The 2D SINFONI flux maps demonstrate that the \ion{H}{I} emission in Th~28 probably consists of two components. In addition to the bright Pa$\beta$ and Br$\gamma$ emission regions with roundish shapes, there is also a faint component stretching along the jet axis. As in the case of other YSOs, the emitting region of atomic hydrogen is very compact. Pa$\beta$ (FWHM=$0\farcs72\pm0\farcs04$) and Br$\gamma$ (FWHM=$0\farcs33\pm0\farcs03$) are barely resolved, with angular sizes similar to those of continuum emission (FWHM$_J$=0\farcs68 and FWHM$_K$=0\farcs32). At the same time, the RVs of the Pa$\beta$ and Br$\gamma$ emission, integrated over the compact bright atomic hydrogen core, are shifted toward the red, although their peak RVs are clearly lower than those of [\ion{Fe}{II}]. The measured redshifts of RV in Pa$\beta$ (34 km s$^{-1}$) and Br$\gamma$ (18 km s$^{-1}$) are similar to those obtained by \citet{Cof10} for the red lobe, and can trace the gas in the descending jet lobe, which is brighter than the blue lobe. Additionally, the redshifted emission can be partly caused by infalling material in the accretion columns. This infalling gas detected in emission lines is assumed for the magnetospheric accretion model, as suggested by observations and modelling of other YSOs \citep{Muz98,Tam16}.

On the other hand, there is a fainter emission component that is spatially resolved and extends along the outflow direction (Fig.~\ref{lines9}). The signs of faint emission are visible in both directions outside the bright \ion{H}{I} emission core. Therefore, both formation scenarios (accretion and jet activity) of atomic hydrogen emission likely contribute to the case of Th~28. However, the spectral resolution of our observations is not high enough to discriminate between these two scenarios. We obtain only the general distribution of RV, and do not resolve any velocity components. The atomic lines have wide emission profiles ($\ga250$~km~s$^{-1}$), which supports the presence of a high-velocity component.

\subsection{Investigating the origin of H$_2$ emission}   

Many collimated young stellar outflows and jets and associated HH objects exhibit molecular hydrogen emission. There are several explanations for the presence of extended H$_2$ emission in the NIR around YSO with jets. The primary excitation mechanisms are (1) H$_2$ emission excited by shock waves, (2)  thermal excitation by UV and X-ray radiation \citep{Gus08}, (3)  irradiation of fluorescent quiescent H$_2$ due to the IR cascade of non-thermal pumping by L$\alpha$ \citep{Beck19}, and (4)  excitation via ambipolar diffusion \citep{Fra99}. There are various scenarios for the formation of spatially extended H$_2$ emission in different gas structures of T Tauri stars. For example, shocks seem to be more responsible for the excitation of H$_2$ emission that originates in the jet lobes \citep{Beck19}. Several scenarios are assumed for the origin of the wide-angle H$_2$ emission, which can originate in a circumstellar disc of a YSO \citep[see Fig.~10 in ][]{Agra14}.
 
The first detections of the $\nu = 1-0\ S(1)$ ro-vibrational transition of H$_2$ at 2.12 $\mu$m in the spectrum of HH objects were identified as shock-excited emission \citep{Bec78,Eli80,Bur89,Wil90}. \citet{Beck08} listed some highly collimated T~Tauri jets revealing shock-excited H$_2$ emission in their jet lobes. \citet{Car15} studied 18 massive jets driven by intermediate-mass and high-mass YSOs and found that all the flows analysed show H$_2$ emission lines that originate from shocks at high temperatures and densities. A recent high-resolution study of massive YSO \object{G192.16-3.82} also detected H$_2$ emission in a chain of tightly collimated knots, implying its shock origin \citep{Bol19}.

However, \citet{Bary99} detected molecular hydrogen gas around CTTS \object{TW~Hya}, which appears to be concentrated in its circumstellar disc. They found that this H$_2$ emission is fluorescent in origin, induced by X-ray radiation. Therefore, this fluorescence from quiescent molecular hydrogen gas may be expected in other X-ray bright T~Tauri stars. Furthermore, \citet{Bary03a,Bary03b} discovered this quiescent H$_2$ emission towards several CTTs, and concluded that it is most likely located in a disc orbiting the young star. \citet{Tak04} detected NIR H$_2$ emission around DG Tau, which is of an  age \citep[$\sim$1--2 Myr,][]{Gue07} similar to that of  Th~28, and concluded that it originates from both a fast material around the jet axis and a slow, poorly collimated molecular component.  Furthermore, \citet{Beck08} found nearly symmetric H$_2$ emission orientated perpendicularly to the outflow axis extending up to $\sim$70 au. 
        
\citet{Beck19} argued that different mechanisms of H$_2$ excitation can coexist in young stellar systems. Therefore, we need tools to determine which are responsible or dominant for the observed H$_2$ emission. In several previous studies, the measured $\nu = 2-1\ S(1)/\nu = 1-0\ S(1)$ line ratio has been used to distinguish between shock excitation and nonthermal excitation processes, such as UV pumping and fluorescence \citep{Gre10,Beck08,Beck19}. Typically, non-thermally excited H$_2$ emission regions have line ratios $\nu = 2-1\ S(1)/\nu = 1-0\ S(1)$ of around 0.55 \citep[UV excitation, ][]{Bla87}, whereas the line ratios arising from shocks are between 0.05 (C-type shocks) and $\sim$0.24 (J-type shocks) \citep{Smith95}. The spectrum of Th~28 shows that the $\nu = 2-1\ S(1)$ line is quite faint, and its emission is concentrated on the stellar source. Therefore, our measurement of the ratio is related to this compact spatial region. The pixel-to-pixel ratio of these two lines, dereddened using $A_V=6$ mag, is $\sim0.03-0.09$ (with uncertainties between $0.007-0.04$), which agrees with those found for several classical T~Tauri star outflows \citep{Beck08}. This value is in accordance with C-type shocks.

At the same time, other ratios of H$_2$ lines can also be used for this goal. For example, the ratio $\nu = 1-0\ S(1)/\nu = 1-0\ Q(1)$ \citep{Beck08} is well determined in our data at positions away from the driving source, and therefore we can compute the ratio for the faint extended V-shaped emission located in the peripheral region. This ratio is predicted to be $\sim$1 in the case of UV-pumped fluorescent H$_2$ \citep{Bla87}, while the line ratios arising from C- and J-type shocks are 1.29 and 1.59, respectively \citep{Smith95,Beck08}. Our measured pixel-to-pixel values of $\nu = 1-0\ S(1)/\nu = 1-0\ Q(1)$, where these lines have  S/N $>3$, vary between a maximum value of $1.23\pm0.08$ and a minimum value at $0.40\pm0.09$, with a mean value of $0.70\pm0.08$, which is lower than predicted by shock excitation and even by UV excitation.

The Th~28 H$_{2}$ emission can be compared with that resolved around \object{T~Tau N}, studied by \citet{Gus08}. In some respects, Th~28 is similar to T~Tau~N, which is also a YSO with spectral type K0, an age of $\sim$1--2 Myr, and a mass of $\sim1 M_\odot$. The outer radius of the spatially resolved H$_2$ disc-like emission is $\sim$ 100 au, which is greater than $R\sim50$ au predicted by the models of \citet{Nom05} and \citet{Nom07}. However, the H$_2$ emitting region in the Th~28 disc is $\sim$2.5 larger than that of T~Tau~N \citep[the Th~28 H$_2$ emitting disc is also $\sim$2.5 larger than its CO-disc measured by ALMA,][]{Lou16}. For the H$_{2}$ emission in T~Tau~N, $\nu = 1-0\ S(1)/\nu = 1-0\ Q(1)=0.9$, which is similar to the case of Th~28. \citet{Gus08} argue that the most likely scenario of extended T~Tau~N H$_{2}$ emission is a wide-angle wind that impinges on a flared disc. Taking all this into account, this scenario can also be applied to explain the H$_2$ emission in the arc-shaped component of Th~28. However, in this scenario, shock excitation is expected, but the ratio $\nu = 1-0\ S(1)/\nu = 1-0\ Q(1)$ is lower than predicted by shocks for both   Th~28 and T~Tau~N. \citet{Gus08} also considered the possibility that H$_2$ emission arises from excitation in the walls of an envelope cavity by an oblique wind, but concluded that this requires an unusually high velocity of the oblique outflow and/or a small opening angle of the cavity. This scenario is also difficult to apply to the Th~28 case for the same reasons. 

Another mechanism that can coexist in the H$_2$ disc is the photoevaporation of the disc by far-UV photons and X-rays \citep{Sto99, Gus08}. This energetic radiation will dissociate molecular hydrogen and ionise atomic hydrogen, resulting in a \ion{H}{I}--H$_2$ transition layer, below which the gas is molecular. As a consequence, this will lead to hydrogen recombination lines such as Pa$\beta$. Furthermore, this region will show strong emission from [\ion{Fe}{II}] 1.257, 1.644 $\mu$m, as well as [\ion{S}{II}] 6730 \AA, in high-density regions. This photoevaporation mechanism may power a collimated disc wind \citep{Gus08}. In the case of Th~28 the H$_2$ emission has a bright central core, which coincides spatially with the emission of Pa$\beta$ and Br$\gamma$. The size of the H$_2$ emission core is similar to that of atomic hydrogen (at 2.122~$\mu$m the H$_2$ core has FWHM=0\farcs38, while Br$\gamma$ has FWHM=0\farcs32). Therefore, we cannot rule out the possibility that we spatially resolve the inner parts of the disc, where this excitation mechanism might dominate. After deconvolution by the PSF, the intrinsic FWHM of the bright H$_2$ core is derived to be $\sim$0\farcs2, giving the photoevaporation radius to 18 au (at 185 pc). This radius is three times larger than that determined for T~Tau~N ($\sim$6 au) by \citet{Gus08}. However, according to the models of \citet{Dul07}, the radius of the photoevaporation region induced by far-UV radiation can range from 3--150 au for solar-mass stars. The photoevaporative disc wind models calculated by \citet{Rab22} predict that the majority of H$_2$ 2.122 $\mu$m luminosity will be confined to a radius of 30 au. Therefore, our estimation of the bright H$_2$ core for Th~28 is within these limits. The wind models of \citet{Rab22} also show a low maximum radial velocity of the H$_2$ 2.122 $\mu$m line, $|\upsilon_\mathrm{p}|<6$ km s$^{-1}$, for a low-inclined circumstellar disc (e.g. Th~28) that in general agrees with our measurements of the H$_2$ RV.

Shock-excited H$_2$ emission in YSO jets represents less collimated and more massive molecular outflows with velocities of the order of $\sim$1-30~km~s$^{-1}$, which are believed to consist of shells of ambient gas swept up by the jet bow shock and an ambient slower and wider angle component \citep{Fra14}. For instance, RW~Aur featuring a high collimated jet shows a low-velocity spatially extended arc-shaped H$_2$ emission \citep{Beck08}. Observations of the jets of RW~Aur and DG~Tau \citep{Mel09,Agra11} also confirm that the jet gas shows a clear drop in velocity toward the jet edges, which is in contrast to the classical X-wind model, where the ejection speed is similar at all angles \citep{Sha07,Fra14}.

Our 2D SINFONI direct images do not show any elongation of the H$_2$ emission along the jet. However, the PV diagram of H$_2$ 2.122 $\mu$m (Fig.~\ref{pv_h2_jet}) reveals faint H$_2$ emission distributed along the jet \citep[this diagram is similar to Fig.~2 from][]{Cof10}. PV diagram of another bright H$_2$ line at 2.407 $\mu$m has the same morphology. Therefore, we calculated the ratio $\nu = 1-0\ S(1)/\nu = 1-0\ Q(1)$ from the PV diagrams. We excluded the central part of the emission around the jet source to avoid a contribution from other subsystems. We found $\nu = 1-0\ S(1)/\nu = 1-0\ Q(1)=1.39\pm0.34$ between 0\farcs5 and 1\farcs3 in the red lobe and $1.29\pm0.26$ between 0\farcs5 and 1\farcs4, corresponding roughly to the C-type shock ratio \citep{Smith95}, but the measured ratio errors are quite high in our case. Since this H$_2$ emission also shows a very low RV, it can arise from a low-velocity ambient gas surrounding the central, higher-velocity lobe. On the other hand, this low H$_2$ emission can also trace a lower velocity molecular wind surrounding the inner fast jet, as suggested for DG Tau \citep{Agra14}.

Another scenario is that this morphology of the H$_2$ emitting region is formed by scattering of the emission by a cavity that the jet drilled into the parent circumstellar envelope. This scenario could explain the bright bump in the H$_2$ emission, which coincides with the shorter blueshifted lobe. Since the measured jet inclination is close to the plane of the sky ($i\sim$80$\degr$), we should see that the circumstellar disc is almost edge-on. YSOs where outflows have swept out most of the mass of the parent cloud can form cavities with biconical morphology, which has been considered for Herbig Ae/Be stars \citep{Fue02}. For instance, \object{LkH$\alpha$~233}, which has a biconical nebulosity, has an age of $\sim$7 Myr \citep{Her04}.

Th~28 is optically visible despite the high inclination of its circumstellar disc. This, as well as its spectral energy distribution, argues against the presence of a massive envelope with bipolar cavities, similar to those of Class 0 and I objects, for which scattering of line and continuum emission has been observed \citep[e.g.][]{Fed20}. Although a tenuous dusty envelope may be present around Th~28, it should scatter not only H$_2$ photons but also those of the continuum. To verify its presence, the $K$-band data cube was spectrally collapsed, after subtracting the line emission. This produced a continuum image with an S/N that exceeded that of the integrated H$_2$ emission image. Comparison with the latter showed that the continuum map has a round and symmetric shape and completely lacks the extended morphology seen in H$_2$. Thus, we conclude that the structure of the H$_2$ emission cannot arise from the walls of the outflow cavity or from dust scattering.

Finally, we conclude that the H$_2$ emission in Th~28 consists of three components: the arc-like extended emission concentrated in the circumstellar disc plane, the bright inner emission centred around the jet source, and the faint bipolar component elongated along the jet lobes. We suggest the following scenarios for the formation of the components. The extended H$_2$ emission in the disc plane is shock-excited by a wide-angle wind that impinges on the flared disc \citep{Gus08}. The origin of this wide-angle component is still debated, but its existence is predicted by models \citep[e.g][]{Cab99}. \citet{Tak06} describe it as an ``unseen wide-angle wind''. Its origin may be the wide-angle molecular wind surrounding the inner fast jet \citep[e.g.][]{Tak04}. The central bright emission core within $R\sim18$ au may represent the region in the disk where the photoevaporation mechanism dominates. The faint bipolar H$_2$ emission probably traces a low-velocity and wide-angle molecular wind surrounding the inner fast and collimated jet.

\subsection{Spectroastrometry of emission photocentre}

The high-spatial-resolution SINFONI images allowed us to reveal positional shifts of the emission photocentre with respect to the position of the stellar continuum as a function of wavelength. A shift in the emission peak was detected in bright lines that belong to different gaseous structures. The strongest shift was discovered in the direction of the jet axis, but the [\ion{Fe}{II}] and H$_2$ also reveal an apparent shift in the orthogonal direction, along the disc axis. The lines with the strongest positional shift are the [\ion{Fe}{II}] lines, where these shifts reach $\sim$30--55 au in the redshifted jet lobe. A smaller shift of $\sim$13~au was detected in the Br$\gamma$ line, as well as H$_2$ 2.413~$\mu$m. This value is more than an order of magnitude lower than that measured using the same line for two massive YSOs \citep{Grave07}. This implies that the line displacement may depend on the mass of the central source.

This emission displacement indicates that we probably resolved the region where the jet is forming and that the [\ion{Fe}{II}] emission reaches its maximum at a greater distance from the central source. Moreover, the same effect for the H$_2$ emission, which arises in an orthogonal plane, implies that these processes can also influence the circumstellar gaseous regions. The size of the H$_2$ positional shift is also smaller than that of [\ion{Fe}{II}] and \ion{H}{I}, which probably means that the surface where H$_2$ interacts with the jet is relatively small. However, the molecular hydrogen emission, as well as the [\ion{Fe}{II}] emission, shows a shift along the circumstellar disc plane, whereas the atomic hydrogen lines do not reveal this behaviour. Together, these details indicate a complex picture of gas interaction in the compact region where the jet forms. 

As in the case of brightness asymmetry, we can also see here a prominent asymmetry in the [\ion{Fe}{II}] photocentre shift found near the central source for opposite jet lobes. The largest shift along the jet axis is detected for the (longer) red lobe, whereas for the (shorter) blue lobe the shift is much smaller. The same displacement can also be observed for the Pa$\beta$ and Br$\gamma$ lines. One of the possible origins could be an occultation effect from the circumstellar disc, which blocks radiation from the jet lobes close to the jet source. Since the circumstellar extinction in the direction to Th~28 is low (see Sect.~\ref{orig_fe}) and the jet axis is close to the sky plane (inclination of 10-15$\degr$), the occultation effect cannot be responsible for the displacement of 55 au in the [\ion{Fe}{II}] 1.644 line.  On the other hand, the [\ion{Fe}{II}] sequence in Fig.~\ref{Feii_14} shows that the absolute flux maximum shifts in either the blue or red direction. At the same time, the absolute flux in the red direction is stronger than in the blue direction (Sect.~\ref{orig_fe}), which is indicative of a real emission shift (or scattering) rather than an occultation effect. This strongly suggests that the formation of the observed brightness asymmetry of the jet lobes may be linked to a different intensity at the level of the jet formation process (such as different mass loss rates in opposite directions) near the driving source. However, the observations presented here cannot distinguish whether this shift belongs to the circumstellar atomic gas or to the gas component that flows in the jet.

\section{Conclusion}
\label{Conc}

 SINFONI high-resolution images reveal the clear and simple structure of the highly collimated bipolar jet of Th~28. The [\ion{Fe}{II}] emission originates in the highly collimated jet lobes, while the bright H$_2$ emission arises from an arc-shaped region orthogonal to the jet axis. Our conclusions can be summarised as follows.

\begin{enumerate}
        \item The arc-like shape of the H$_2$ emission indicates that the excited H$_2$ is concentrated in the circumstellar disc and that this excitation can be caused by shocks from a wide-angle stellar wind impinging on the disc, rather than excited by shocks in the jet or fluorescence. The size of the spatially resolved emission disc, measured from the brightest H$_2$ line at 2.122~$\mu$m, is $\gtrsim$540 au.
        \item In addition to the arc-shaped H$_2$ emission detected directly on the SINFONI images, PV diagrams of the bright H$_2$ lines reveal faint H$_2$ emission along both jet lobes, as also reported by \citet{Cof10}. The H$_2$ PV diagrams show that the molecular hydrogen emission has a very low velocity compared to forbidden line emission, traced by the [\ion{Fe}{II}] lines.
        \item The Pa$\beta$ and Br$\gamma$ lines reveal a two-component morphology. Although the unresolved bright \ion{H}{I} emission is concentrated on the central jet source position, there is also a faint extended emission associated with the jet lobes. The line width of this \ion{H}{I} emission is about $\sim250$~km~s$^{-1}$.
        \item The arc-shaped H$_2$ emission shows a low-velocity pattern. The measured width of the H$_2$ lines is $\lesssim$100~km~s$^{-1}$,  about 2.5 times less than that of Pa$\beta$ and Br$\gamma$.
        \item Two gaseous knots visible in [\ion{Fe}{II}] are detected in the bipolar jet, one in each lobe, at angular distances of 1\arcsec{} in the blue lobe and 1\farcs2 in the red lobe.
        \item The emission centroid in all gas lines is shifted with respect to the central stellar jet source, which indicates that the observations resolve a region where the inner jet forms. Therefore, the [\ion{Fe}{II}] emission displacement traces a change in the excitation of the gas, as the jet propagates in the medium. The \ion{H}{I} emission displacement traces the neutral hydrogen that has been entrained by the jet from the accretion region, or the inner hot parts of the jets which is likely to emit in \ion{H}{I} lines.
        \item The observed extension of the opposing jet lobes, traced by [\ion{Fe}{II}] emission, correlates with the value of the [\ion{Fe}{II}] photocentre shifts with respect to the jet source. This suggests that the observed brightness asymmetry is intrinsic and arises in the immediate vicinity of the driving source of Th~28.
        \item The maximum size of the jet launch region is derived as 0\farcs 015, which corresponds to 3 au, and the initial opening angle of the Th~28 jet is obtained as $\sim$28\degr, making this jet substantially less collimated than most of the jets from other CTTs.
\end{enumerate}

Previous NIR 1D spectroscopy of Th~28 jet provided valuable information on the conditions and morphology of the gas in the jet. The results presented in the current study demonstrate that 2D spectroscopy with the SINFONI IFU in the NIR is an effective tool for the study of the gas morphology at high spatial resolution within a $3\arcsec\times3\arcsec$ field around the central source of Th~28. Spectroscopy of other YSO jets has shown that several optical forbidden doublets ([\ion{O}{I}] 6300,6363, [\ion{N}{II}] 6548,6583, [\ion{S}{II}] 6716,6731 \AA) are usually bright in the jet spectrum. Going beyond the present study, we note that 2D spectroscopy of the forbidden line doublets at optical wavelengths with a similarly high spatial resolution would allow us to directly measure basic parameters of the jet in Th 28 in detail, as has been successfully done for several other YSO jets.

\begin{acknowledgements}
This work is based on observations obtained with ESO VLT/SINFONI (Paranal, Chile) within the observing programme 095.C-0892(A). We thank the referee for comments that helped to improve the paper. A.C.G. has been supported by PRIN-INAF-MAIN-STREAM 2017 ``Protoplanetary disks seen through the eyes of new-generation instruments'' and by PRIN-INAF 2019 ``Spectroscopically tracing the disk dispersal evolution (STRADE)''. RGL acknowledges support from Science Foundation Ireland under Grant No. 18/SIRG/5597. PAB and NSN acknowledge the support of the Russian Science Foundation, grants 18-72-10132 and 21-72-03016, for the initial stages of this project. This research made use of the SIMBAD database, operated at CDS, Strasbourg, France.
\end{acknowledgements}

\bibliographystyle{aa}
\bibliography{paper_jet}

\end{document}